\documentclass[final,authoryear,5p,times,twocolumn]{elsarticle}

\usepackage{amssymb}
\usepackage{amsmath}

\usepackage{enumerate}
\usepackage{times}
\usepackage{jtb}

\DeclareMathAlphabet{\bi}{OML}{cmm}{b}{it}

\def\gmas{\gamma_{+}}
\def\gmenos{\gamma_{-}}

\journal{Journal of Theoretical Biology}

\begin{document}

\begin{frontmatter}



\title{Species assembly in model ecosystems, I: Analysis of the population model and the invasion dynamics.}


\author{Jos\'e A.\ Capit\'an}
\ead{jcapitan@math.uc3m.es}
\author{Jos\'e A.\ Cuesta\corref{cor}}
\cortext[cor]{Corresponding author.} \ead{cuesta@math.uc3m.es}
\address{Grupo Interdisciplinar de Sistemas Complejos (GISC), 
Departamento de Matem\'aticas, Escuela Polit\'ecnica Superior, 
Universidad Carlos III de Madrid, E-28911 Legan\'es, Madrid, Spain}

\begin{abstract}
Recently we have introduced a simplified model of ecosystem 
assembly \citep{capitan:2009} for which we are able to map out
\emph{all} assembly pathways generated by external invasions in an \emph{exact}
manner. In this paper we provide a deeper analysis of the model, obtaining
analytical results and introducing some approximations which allow us to
reconstruct the results of 
our previous work. In particular, we show that the population dynamics
equations of a very general class of trophic-level
structured food-web have an unique interior equilibrium point which is globally
stable. We show analytically that communities found as end states
of the assembly process are pyramidal and we find that the equilibrium
abundance of any species at any trophic level is approximately inversely proportional 
to the number of species in that level. We also find that 
the per capita growth rate of a top predator
invading a resident community is key to understand the appearance of complex
end states reported in our previous work. The sign of these rates 
allows us to separate regions in the space of parameters where the end state
is either a single community or a complex set containing more than one
community. We have also built up analytical
approximations to the time evolution of species abundances that allow us to 
determine, with high accuracy, the sequence of extinctions that an invasion may cause.
Finally we apply this analysis to obtain the communities in the end states.
To test the accuracy of the transition probability matrix generated
by this analytical procedure for the end states, we have compared averages 
over those sets with those obtained from the 
graph derived by numerical integration of the Lotka-Volterra equations.
The agreement is excellent.
\end{abstract}


\end{frontmatter}

\section{Introduction}
\label{intro}

A piece of common wisdom in ecology is that biodiversity enhances the stability
of ecosystems. This has traditionally been a well established observational fact since
the works of \cite{odum:1953}, \cite{macarthur:1955} and
\cite{elton:1958} who showed that simple ecosystems (e.g.\ man-cultivated lands)
undergo very large fluctuations in population and are vulnerable to invasion,
an effect that gets reduced upon increasing the number of predators and
preys in the system. But early in the 70s May showed
that randomly generated dynamical models for the populations of a community exhibit
the opposite feature: the larger the species abundance the smaller its linear stability
\citep{may:1972,may:1973}. Thanks to this controversy we have
gained very much insight into the nature of ecosystems \citep{mccann:2000}.
Apart from the introduction of more refined concepts of ecosystem stability
\citep{pimm:1982}, one
of the main conclusions arising from the comparison of empirical data with
May's predictions on the bounds for community stability \citep{dunne:2006} is
that real ecosystem are within the tiny set of stable ones, no matter
how large they are; in other words, ecosystems are far from being just random
gatherings of species.

Natural communities carry out a selection mechanism that induces colonizers
adaptation. There has been a lot of theoretical work in the past devoted to study
the assembly of communities through successional invasions
\citep{post:1983,drake:1990,case:1990,law:1993,law:1996,morton:1997}.
Overall, these papers have provided a
theoretical framework to understand how communities are built up \citep{law:1999}.
The basic process in which these models are based is
the sequential arrival of rare species (invaders) that colonize the ecosystem
and that may be established, possibly causing a global reconfiguration of the
community in the long term by means of several species extinctions. 
Obviously, these models are but idealizations
of the complex processes taking place in real community assembly, but
simple mechanisms acting in these models could be expected to be the ones 
responsible for the formation of real ecosystems \citep{law:1999}. This approach 
of devising theoretical paradigms for real situations 
has been successfully applied over and over in the field of statistical
mechanics ---where, for instance, using such an idealization as the Ising model
provides the clues to understanding ferromagnetism in real materials \citep{huang:1987}.

Previous assembly models tend all to rely on the Lotka-Volterra dynamics
[but see the recent work of \cite{lewis:2007}], although
differ in the criterion to accept an invasion. While \cite{post:1983}
assumed that new species were created \emph{ad hoc,} according to certain stochastic
rules, subsequent approaches \citep{drake:1990,law:1996} introduced the concept of
``species pool''. A regional species pool is a set of possible invaders whose trophic
interactions have been determined in advance  \citep{law:1996}. Despite these
differences, all previous papers arrive at the conclusion that the species richness
of each resident community increases along successional time, although the
average resistance of a community to be colonized increases in time. Therefore community
assembly increases biodiversity as well as stability, understood as resistance to invasions. 

Nevertheless, one must bear in mind that not all assembly pathways
have been explored in these models. The conclusions reached so far
rely on averages of quantities under
study over a finite set of realizations of the underlying stochastic process, that is ultimately
based on a {\em finite} pool of possible invaders. This has raised several question that
remained without a definitive answer. For example, there was no clear-cut answer
regarding the dependence of the results on the history of invasions.
\cite{morton:1997} found a final
end state resistant to invasions by the remaining species in the pool at the end of the process,
and this end state could be either a single ecosystem or a set involving more than one
community connected by invasions with one another. Despite this conclusion,
the dependence of the end state on the assembly history is a matter of discussion
\citep{fukami:2003}. Moreover, we should not forget that the number of species in the pools
employed is always relatively small, so the question remains as to whether larger pools
lead to qualitatively different results. In this respect, it has been pointed out
\citep{case:1991,levine:1999} that the exhaustion of good invaders in the early assembly
might be just an artifact of the finiteness of the pool.

Trying to overcome the shortcomings of previous models, in our previous work \cite{capitan:2009} 
we proposed a minimalistic model of ecosystem assembly 
with which we were able to analyze \emph{all}
assembly pathways, thus characterizing the full assembly process. In spite of its
simplicity, we recovered the same conclusions found previously.
Our model is also based on a pool of species and a niche variable (the trophic level) 
that determines their interactions. In contrast, however, our pool is infinite. 
In spite of
that, within the assumptions of the model, we found a finite number of
(viable) communities linked by colonization. This allowed us to define an assembly graph 
for our model ---similar to that of \cite{warren:2003}, who studied the assembly process
experimentally for a small pool of 6 protist species. By assigning
transition probabilities to the links of this graph the assembly process
was mapped to a Markov chain \citep{karlin:1975}, which is tantamount to saying that we
defined a statistical mechanics on the set of viable communities (microstates). In other
words, our model gives the probability distribution of all these
microstates at any time. This allowed us to characterize both transient and
equilibrium states, as well as to compute the time evolution of any observable
along the assembly in an exact manner. But more importantly, as our model provides a complete
and exact (albeit numeric) description of the assembly process, we can positively state
that, under the assumptions of our model, in the long-term assembly dynamics a unique
enstate is reached, and this state is formed by just one uninvadable community or a
closed set of communities connected between them. These sets contain the communities
that survive in the long term, and the ecosystem can be regarded
 as a fluctuating community that can vary each
level occupancy trough successional invasions. 

In this paper we will give some analytical results for the underlying
population dynamics of our assembly model, and we will see how these results can be 
combined together to arrive at the same conclusions we obtained numerically
in our previous work. Relying on these analytic results, we
will be able to describe the observables that characterize
the end states with high accuracy. 
In particular, we will reproduce the variation of the number of communities
in each end state with the abundance of abiotic resources, as well as the average values
of quantities like the species richness. We will leave the computational and numerical 
results that can be obtained with this model for the second paper of this suite 
\citep{capitan:2010b}, which will be focused in the successional variation of biologically
relevant quantities along the assembly, and the analysis of the main properties of transient
states. 

This paper is organized as follows. Section \ref{s:trophic} is devoted to the analysis
of a rather general model of trophic-level structured food-webs, and the discussion
of its dynamic stability. In Section \ref{s:symmetry} we will restrict ourselves to
a particular case of community by making a species symmetry assumption, that renders
our model closer to neutral models and allows a more detailed analytical study. In 
Section \ref{s:rest} we will deduce some analytical properties of the equilibrium 
point, such as estimations of the maximum number of species
allowed in a community for a given set of parameters, or the maximum number of
trophic levels that the amount of resource allows. Section \ref{s:invaded} is dedicated to
discussing some criteria for an invader to establish in a community, and to give
some global analytical approximations to the time evolution of a system invaded by
a top predator. Finally, in Section \ref{s:assembly} we will apply our analysis
to recover the results obtained in \cite{capitan:2009}
by means of a numerical integration of the population dynamics equations.

The two papers of this suite are self-contained and can be read separately, although
they are cross-referenced. Readers interested in the underlying population dynamics
of our model will find a detailed discussion in this paper. Those
readers more interested in the ecological consequences and results that the model
provides can skip the technical Sections \ref{s:rest} and \ref{s:invaded}.
For a full account of the results that we have obtained, we refer them to the 
companion paper.

\section{Trophic-level structured food-webs}
\label{s:trophic}

How species are arranged in a network to conform a food-web is a question difficult 
to answer. The specific topology of the network
where feeding interactions take place is very complex  and several complicated models have been 
proposed for both the structure and the dynamics of food-webs \citep{dunne:2006}. 
In contrast, our aim in \cite{capitan:2009} was to construct a minimalistic model, so  
we considered the traditional picture of trophic pyramids of interacting species in 
different, well defined trophic levels. Although
trophic levels can be roughly described in real webs \citep{martinez:2006}, we will assume that 
feeding interactions take place strictly between species belonging to contiguous, 
well defined trophic 
levels. This is a standard (and accurate) assumption, as the models of tri-trophic
food chains show \citep{bascompte:2005}.
This notwithstanding, it is acknowledged that {\em omnivory}, i.e. predation from 
several levels, exists although is still an open question how common it is. For example,
work on food-web motifs has found that omnivory is sometimes under-represented and
sometimes over-represented in real networks \citep{bascompte:2005}. However, the
impact of including omnivory in the model could lead to non trivial results. Since the
trophic level is normally related to species size, feeding from lower levels will provide
less energy to predators, so proper allometric relations should be included in the model
to fix the interaction strengths. For the sake of simplicity, we will not divert ourselves 
from the standard assumption of disregarding omnivory.

Therefore, any species at level $\ell$ will feed only on 
species at level $\ell-1$ and will be predated only by species at level $\ell+1$. Let
$s_{\ell}$ be the number of species in the $\ell$-th level. Thus for an ecological
community with $L$ trophic levels the total number of species is 
$S=\sum_{\ell=1}^L s_{\ell}$. In order to determine which species are 
predated at each level, we define the set of 
interaction matrices $\Gamma^{\ell}$, with
dimensions $s_{\ell} \times s_{\ell-1}$, such that the element $\Gamma_{ij}^{\ell} = 1$
when species $j$ in level $\ell-1$ is a prey of species $i$ in level $\ell$, and is
zero otherwise. Any particular choice of this set of matrices
determines the food-web in our model.

According to our aim of developing a simplified model, we propose a simple population
dynamics with the purpose of capturing on average the main behavior of species
abundances. It is inspired in a model used before to study coexistence in
competing communities \citep{lassig:2001,bastolla:2005a,bastolla:2005b}. Population 
dynamics is modelled by Lotka-Volterra equations, including both
predator-prey interactions as well as intra- and interspecific competition.
Thus, in order to keep the model minimalistic we have chosen not to include other
interaction types such as mutualism.

Let $n^{\ell}$ be a column vector with the population densities of all the species
at trophic level $\ell$. Following \cite{bastolla:2005a} we propose the mean-field dynamics
\begin{equation}\label{eq:L-V}
\frac{\dot{n}_i^{\ell}}{n_i^{\ell}} = \left(-\alpha+\gmas^{\ell} \Gamma^{(\ell)}n^{\ell-1}
-B^{\ell} n^{\ell} -\gmenos^{\ell} (\Gamma^{\ell+1})^{\rm T} n^{\ell+1}\right)_i.
\end{equation}
We assume that the strength of the feeding interactions between contiguous levels is 
fixed and determined by the constants $\gmas^{\ell}$,
which control the amount of energy available to reproduction for each predation event for 
species at level $\ell$, and $\gmenos^{\ell}$ $(>\gmas^{\ell})$, which take into account the 
mean damage 
caused by predation over level $\ell$. The ratio $\gmas^{\ell}/\gmenos^{\ell}$ measures
the efficiency of conversion of prey biomass into predator biomass.

Interspecific competition in a trophic level is measured by the
off-diagonal elements of the $s_{\ell}\times s_{\ell}$ 
matrix $B^{\ell}$, while intraspecific competition (diagonal 
elements) is normalized to unity (this just amounts to 
fixing a time scale for the dynamics). A natural way to represent this matrix is
\begin{equation}
B^{\ell} = (1-\rho^{\ell})\mathbb{I} + \rho^{\ell} K^{\ell},
\end{equation}
where $\rho^{\ell} \leq 1$ measures the relative magnitude between intra-- and interspecific
competition, and $\mathbb{I}$ is the identity matrix. 
Diagonal elements of $K^{\ell}$ are equal to 1 
due to the normalization of the intraspecific competition.
We will assume (the reasons will become clearer later) that the competition
matrix is symmetric and positive definite. 

Interspecific competition due to sharing common preys is implicitly represented in the predation
terms. There is however a direct competition due to other effects, such as territorial competition,
mutual aggressions, etc. We will assume [as in \cite{bastolla:2005b}] that species
sharing more preys are closely related ecologically 
[this fact might have support from a evolutionary viewpoint
as shown in \cite{rezende:2007}], so their requirements are similar
and we can assume that elements of $K^{\ell}$ are proportional to the
ecological overlapping between species \citep{lassig:2001,bastolla:2005b}. Let
$\pi_{ij}^{\ell}$ represent the number of common preys for species $i$ and $j$
belonging to level $\ell$. The species overlapping due to common preys is 
$K_{ij}^{\ell}=\pi_{ij}^{\ell}/\sqrt{\pi_i^{\ell}\pi_j^{\ell}}$, with $\pi_i^{\ell}$
the total number of preys of species $i$. Under our matrix notation,
$\pi_{ij}^{\ell}=(\Gamma^{\ell}{\Gamma^{\ell}}^{\rm T})_{ij}$ and $\pi_i^{\ell}
=(\Gamma^{\ell}{\Gamma^{\ell}}^{\rm T})_{ii}$, so that
\begin{equation}\label{eq:competition}
B^{\ell} = (1-\rho^{\ell})\mathbb{I} +
\rho^{\ell}D^{\ell}\Gamma^{\ell}(D^{\ell}\Gamma^{\ell})^{\rm T},
\end{equation}
$D^{\ell}$ being a diagonal matrix with elements 
$(\Gamma^{\ell}{\Gamma^{\ell}}^{\rm T})_{ii}^{-1/2}$.
Expressed as \eqref{eq:competition}, it is evident that such a competition matrix is
symmetric and positive definite. It is worth mentioning that this system
does not fulfil the hypotheses leading to Gause's competitive exclusion principle
\citep{hofbauer:1998,bastolla:2005a}, even when there is a single level. Among other
things, this is due to the fact that competition coefficients between different species
are not all the same. This point will be discussed in more detail in the second
paper of this suite \citep{capitan:2010b}.

We regard all the species as consumers,
and so they have a death rate, $\alpha_i^{\ell}$, which is the $i$-th component of 
vector $\alpha^{\ell}$. 
Note that in a real food-web the interaction coefficients will
not be uniform within a trophic level. In this sense, we represent interactions averaged
(mean-field) in each level but we allow variation in the strength of the interactions 
among different trophic levels. Finally, all species at the first level predate on a single 
resource, whose time evolution is given by
\begin{equation}\label{eq:resource}
\frac{\dot{n}^0}{n^0} = R-n^0-\gmenos^1 (\Gamma^1)^{\rm T} n^1.
\end{equation}
The constant $R$ is the maximum amount of resource in the absence of its consumers. The 
abundance $n^0$ has to be understood as the amount of a primary abiotic resource, like 
sunlight, water, nitrogen, etc.
It has to be considered as an energetic input for the maintenance of the remaining species 
in the community.
The amount of such resource is limited, hence the saturation of $n^0 $ at a value $R$.

The model is supplemented by an extinction threshold, $n_c>0$, independent of the
species. If a population falls below this value it is considered extinct (real
populations can not be arbitrarily small). This viability condition
has been previously used in similar models \citep{kokkoris:1999,borrvall:2000,eklof:2006}, and
accounts for the vulnerability of low density communities against external
environmental variations or adverse mutations \citep{pimm:1991}. The technical need for this
extinction threshold in our model will become clearer when we describe the variation of
the densities in terms of the occupancy of each level.

\subsection{Dynamic stability of the interior equilibrium point}
\label{ss:dynstab}

Equations \eqref{eq:L-V}, \eqref{eq:resource}
have several equilibria. Among them, the main one is obtained
by equating the right-hand side of these equations to zero. If all the equilibrium densities
are positive, this fixed point is called the interior equilibrium. 
The population $p^{\ell}$ at equilibrium 
are obtained as the solution of the linear system of $S+1$ equations
\begin{equation}\label{eq:interior}
\begin{split}
&\gmas^{\ell}\Gamma^{\ell}p^{\ell-1}-B^{\ell}p^{\ell}-
\gmenos^{\ell+1}(\Gamma^{\ell+1})^{\rm T}p^{\ell+1}=\alpha^{\ell}, \\
&p^0+\gmenos^1 (\Gamma^1)^{\rm T}p^1=R.
\end{split}
\end{equation}
for $\ell=1,\dots,L$.
The remaining equilibria are obtained by setting to zero any subset
of the populations and solving the linear system resulting from eliminating those
variables. The resulting system is the same as \eqref{eq:interior}
but if species $i$ at level $\ell$ has zero equilibrium abundance, the $i$-th
column in the corresponding matrix $\Gamma^{\ell}$ has to be eliminated. 
Therefore one only needs the solutions of the linear systems \eqref{eq:interior} for
a given choice of the set of matrices $\{\Gamma^{\ell}\}_{\ell=1}^L$ 
in order to fully determine all the equilibrium densities.

Since feeding relations are established among contiguous levels, 
\eqref{eq:interior} acquires a block-tridiagonal structure. 
Due to this form, the interior equilibrium can be 
formally obtained by applying Gaussian elimination. We put the equilibrium abundances 
in the form 
\begin{equation}\label{eq:thomas}
p^{\ell-1} = M^{\ell} p^{\ell} + c^{\ell}
\end{equation} 
for certain $s_{\ell-1}\times s_{\ell}$ matrices $M^{\ell}$ and $s_{\ell-1}\times 1$ 
vectors $c^{\ell}$ to be determined ($\ell = 1,\dots, L+1$).
Substitution into \eqref{eq:interior} gives the following recursive relations
for $M^{\ell}$ and $c^{\ell}$,
\begin{equation}\label{eq:Mc}
\begin{split}
M^{\ell+1} &= \gmenos^{\ell+1}\left(\gmas^{\ell}\Gamma^{\ell}M^{\ell}-B^{\ell}\right)^{-1}
(\Gamma^{\ell+1})^{\rm T}, \\
c^{\ell+1} &= \left(\gmas^{\ell}\Gamma^{\ell}M^{\ell}-B^{\ell}\right)^{-1}
\left(\alpha^{\ell}-\gmas^{\ell}\Gamma^{\ell}c^{\ell}\right).
\end{split}
\end{equation} 
Since the resource can only be predated and there is no competition, 
we set $\Gamma^0 = 0$ and $\rho^0=0$. This leads to the initial 
conditions $M^1 = -\gmenos^1 (\Gamma^1)^{\rm T}$ and $c^1 = -R$ according to \eqref{eq:resource}. 
Thus, given a particular set of matrices $\{\Gamma^{\ell}\}_{\ell=1}^L$, \eqref{eq:Mc} 
fully determines $M^{\ell}$ and $c^{\ell}$. After that, 
starting from the boundary condition $p^{L+1}=0$ (the community has 
exactly $L$ trophic levels), 
we backsubstitute in \eqref{eq:thomas} to get the equilibrium densities.

We can push further the property that our dynamical system \eqref{eq:L-V} is 
block-tridiagonal to study its dynamic stability.
Let us show that interior equilibria $p_i^{\ell}$, for all $i=1,\dots,s_{\ell}$ and 
$\ell=0,\dots,L$, are globally stable. This result is based in the existence a Lyapunov 
function \citep{hofbauer:1998}, which guarantees that any positive initial condition
evolves towards the interior 
equilibrium. The Lyapunov function for this system is 
\begin{equation}\label{eq:lyapunov}
\mathcal{V}(\{n^{\ell}\}) = \sum_{\ell=0}^L A_{\ell}
\sum_{j=1}^{s_{\ell}} \left(n_j^{\ell}-p_j^{\ell}\log n_j^{\ell}\right)
\end{equation}
where $A_k = \prod_{\ell=1}^{k} \frac{\gmenos^{\ell}}{\gmas^{\ell}}$
for $k = 1,\dots, L$ and $A_0 = 1$. 

For \eqref{eq:lyapunov} to be a Lyapunov function, we just need to check that 
$\dot{\mathcal{V}}\leq 0$ 
along any orbit $\{n^{\ell}(t)\}_{\ell=0}^L$ starting with positive initial abundances 
\citep{hofbauer:1998}. Let us compute its time derivative. If we consider the displaced 
variables
\begin{equation}
y_j^{\ell} = n_j^{\ell}-p_j^{\ell},
\end{equation}
we can write \eqref{eq:L-V} as $\dot{n}_i^{\ell} = n_i^{\ell}q_i^{\ell}$, where
\begin{equation}\label{eq:ql}
q^{\ell} = \gmas^{\ell}\Gamma^{\ell}y^{\ell-1}-B^{\ell}y^{\ell}
-\gmenos^{\ell}\left(\Gamma^{\ell+1}\right)^{\rm T}y^{\ell+1},
\end{equation}
hence the time derivative is simply 
$\dot{\mathcal{V}}(\{n^{\ell}\}) = 
\sum_{\ell=0}^LA_{\ell}\sum_{j=1}^{s_{\ell}} y_j^{\ell}\, q_j^{\ell}$.
After substituting \eqref{eq:ql}, we arrive at
\begin{equation}
\begin{split}
\dot{\mathcal{V}}(\{n^{\ell}\}) = &-\sum_{\ell=0}^L 
A_{\ell}(y^{\ell})^{\rm T}B^{\ell} y^{\ell}\\
&+\sum_{\ell=0}^{L-1} \left(A_{\ell+1}\gmas^{\ell+1}-A_{\ell}\gmenos^{\ell+1}\right)
(y^{\ell+1})^{\rm T}\Gamma^{\ell+1}y^{\ell}.
\end{split}
\end{equation}
Thus our previous choice of $A_k$ cancels the second sum. Since $B^{\ell}$ 
is positive definite, we deduce that the time derivative of the 
Lyapunov function is negative along any orbit, 
and therefore Lyapunov's theorem \citep{hofbauer:1998} ensures the global stability of the
non-trivial rest point $p^{\ell}$. Note that the existence of this Lyapunov function 
is a direct consequence of the block-tridiagonal structure of 
the dynamical system \eqref{eq:L-V}--\eqref{eq:resource}, hence the assumption of
predation only between contiguous levels ensures this global stability.


\section{Species symmetry assumption}
\label{s:symmetry}

In what follows, we will restrict ourselves to the dynamical system \eqref{eq:L-V} with the 
particular choice of interaction matrices $\Gamma^{\ell}_{ij} = 1$ 
for any $i,j,\ell$. This was the system
studied in \cite{capitan:2009}. This assumption implies that all species are
generalist, and the model can now be regarded as a mean-field-like picture of  
real communities, since all species in contiguous levels
interact with each other. We will assume as well 
that interaction coefficients are independent of the trophic level, and we will simply
denote them as $\gmas$, $\gmenos$, $\rho$ and $\alpha$. 
These parameters should now be understood as an average strength of the processes 
involved in the population dynamics. 
These kind of models, which do not make any explicit difference among species, are referred to 
as neutral \citep{hubbell:2001,etienne:2007}. From the point of view of the 
trophic interactions there is no difference between species (neither the rates 
nor the set of preys they feed on make any distinction 
among species). We introduce this symmetric scenario because it will allow a simpler,
analytical description of the community.

Pure neutral models do not make any distinction whatsoever
between species. This is not our case, 
because species can be distinguished by their different balance between 
intra-- and interspecific competition. Neutrality
in our model has to be understood as a species symmetry assumption \citep{alonso:2008} 
for the strength of the interactions. We will discuss  
the case $\rho=1$, when the model turns to be fully symmetric (strictly neutral), 
in the second paper of this suite \citep{capitan:2010b}. 

Under this symmetry assumption, the population dynamics \eqref{eq:L-V} with the competition 
matrix \eqref{eq:competition} transforms into $\dot{n}_i^{\ell} = q_i^{\ell} n_i^{\ell}$, 
where
\begin{equation}\label{eq:L-Vsym}
\begin{split}
q_i^{\ell} &= -\alpha+\gamma_{+}N^{\ell-1} - (1-\rho)n_i^{\ell} - \rho N^{\ell} 
-\gamma_{-}N^{\ell+1}, \\
q^0 &= R-n^0-\gmenos N^1,
\end{split}
\end{equation}
being $N^{\ell} \equiv \sum_{i=1}^{s_{\ell}} n_i^{\ell}$. The set of equations 
\eqref{eq:interior} for the interior rest point imply that the equilibrium abundances 
are equal for any two species $i$ and $j$ of the same level. Hence the equilibrium abundances 
$\{p^{\ell}\}_{\ell=1}^L$ are the solution to the linear system
\begin{equation}\label{eq:intsym}
\begin{split}
\alpha &= \gmas s_{\ell-1}p^{\ell-1} -[1+\rho(s_{\ell}-1)]p^{\ell} 
-\gmenos s_{\ell+1}p^{\ell+1}, \\
R &= p^0+\gmenos s_1p^1,
\end{split}
\end{equation}
for $\ell=1,\dots,L$. Note that the global stability result holds only for this equilibrium 
point.

\subsection{Reduced dynamical system}
\label{ss:reduced}

As in our previous work \citep{capitan:2009}, equilibrium communities will undergo invasions. 
Thus we are interested in the time dynamics of an invaded community initially at 
equilibrium. Notice that the per capita growth rates \eqref{eq:L-Vsym} 
satisfy the equality
\begin{equation}\label{eq:sym}
q_i^{\ell}(\dots,n_i^{\ell},\dots,n_j^{\ell},\dots) = 
q_j^{\ell}(\dots,n_j^{\ell},\dots,n_i^{\ell},\dots)
\end{equation}
under the interchange of the abundance of 
two species at the same level. This symmetry, together with an initial
condition where $n_i^{\ell}(0)=n_j^{\ell}(0)$, is enough to show that the time evolution of
both species is identical (see \ref{s:appA}). Thus we can reduce our dynamical 
system to a set of $L+1$ differential equations,
\begin{equation}\label{eq:L-Vred}
\begin{split}
\frac{\dot{n}^{\ell}}{n^{\ell}} &= -\alpha+\gamma_{+}s_{\ell-1}n^{\ell-1} 
- [1+\rho(s_{\ell}-1)]n^{\ell} -\gamma_{-}s_{\ell+1}n^{\ell+1}, \\
\frac{\dot{n}^0}{n^0}  &= R-n^0-\gmenos s_1 n^1.
\end{split}
\end{equation}

There is another crucial difference between our model and usual neutral models in the
literature. Although neutral models ignore species identity, they are stochastic. It is the 
ecological drift what makes species abundances to stochastically vary. This stochasticity is 
the ultimate reason
for extinction in neutral models. On the contrary, our dynamical system is deterministic. 
The reason to include the (somehow arbitrary) extinction threshold $n_c$ is to ``mimic'' this 
fluctuation-driven extinction of species with low abundance. 

Thus extinctions must be understood stochastically in our model. As it was pointed out in 
\cite{capitan:2009}, the stochastic effect of adverse mutations or external variations of
the environment that make species to go extinct is taken into account in our deterministic
dynamics with the viability condition $n^{\ell} \geq n_c$. Notice however that, strictly speaking, 
when a species of one level falls below $n_c$ the whole level does too. 
Extinguishing the whole level as the strict dynamics would require would be unrealistic. Instead
we eliminate species
one by one until viability is recovered \citep{capitan:2009}. This latter dynamics would
approximate better what one would find in a truly stochastic neutral model, in which the 
simultaneous extinction of general species is very unlikely to happen.

\subsection{Structural stability}
\label{ss:structstab}

We have chosen the constants to be uniform in our model, this making all species on
each trophic level at equilibrium have equal abundance. However, 
according to competitive exclusion \citep{macarthur:1964},
a tiny variation in the parameters that makes any difference among species
will make the system unstable. Fortunately, for this class of
models the competitive exclusion principle does not hold as such. This has been discussed at length
in \cite{bastolla:2005a}. In this paper the authors derive some bounds to the variation allowed
for the constants that the system can tolerate without leading any species to extinction. In
fact, the dynamical system they discuss is the same as we have described in Section \ref{s:trophic},
with different constants for different species. The more diverse the ecosystem is the stricter
are these bounds, but in any case, no matter how diverse the ecosystem is, some variation of
the constants is always tolerated without this leading any species to extinction. This proves
the structural stability of our system, even under the assumption of species symmetry.

\section{Analytical properties of the interior rest point}
\label{s:rest}

\subsection{Maximum number of species and maximum number of levels}
\label{ss:maxS}

In this subsection we will obtain an analytical estimation of the maximum number of species 
that a trophic level can host	
among all the possible viable equilibria. We simply set 
all the abundances in each level to be equal to $n_c$ and solve the resulting linear system
\eqref{eq:intsym} for $\{s_{\ell}\}_{\ell=1}^L$ and $s_0\equiv p^0/n_c$,
\begin{equation}\label{eq:rec}
\begin{split}
&s_0 + \gmenos s_1 = \frac{R}{n_c},\\
&\gmas s_{\ell-1} - \rho s_{\ell} - \gmenos s_{\ell+1} = 1-\rho+\frac{\alpha}{n_c},
\end{split}
\end{equation}
for $\ell\geq 1$. We introduce the generating function 
$G(z)=\sum_{\ell=0}^{\infty} s_{\ell} z^{\ell}$ for the sequence $\{s_{\ell}\}_{\ell=1}^L$.
The explicit solution will depend on two initial
conditions $s_0$ and $s_1$, since we have a two-term recursion. We will leave them undetermined
for the moment. The second equation of \eqref{eq:rec} allows us to 
calculate explicitly $G(z)$,
\begin{equation}
G(z)=\frac{(1-\rho+\alpha/n_c)z^2}{(1-z)(\gmas z^2-\rho z-\gmenos)}
-\frac{\gmenos s_0+z(\rho s_0+\gmenos s_1)}{\gmas z^2-\rho z-\gmenos}.
\end{equation}
We recover the general term of $s_{\ell}$ by a series expansion of the 
generating function. Let us first define the constants 
$\mu=(1-\rho+\alpha/n_c)/(\gmenos-\gmas+\rho)$ and 
$z_{\pm}=(\rho\pm\sqrt{\rho^2+4\gmas\gmenos})/(2\gmas)$.
In order to get compact expressions, we define the auxiliary sequence
\begin{equation}\label{eq:al}
a_{\ell}=\left(\frac{\gmas}{\gmenos}\right)^{\ell}\frac{z_+^{\ell+1}-z_-^{\ell+1}}{z_+-z_-},
\end{equation}
which satisfies the two-term recursion $\gmenos a_{\ell}=\rho a_{\ell-1}+\gmas a_{\ell-2}$
with initial conditions $a_{-1}=0$, $a_0=1$. This recurrence can be fully expressed
as a linear combination of powers of $\rho/\gmenos$ and $\gmas/\gmenos$,
\begin{equation}\label{eq:als}
a_{\ell}=\sum_{k=0}^{\lfloor \ell/2 \rfloor}\left(\begin{array}{c}
\ell-k \\
k \end{array}\right)\left(\frac{\rho}{\gmenos}\right)^{\ell-2k}
\left(\frac{\gmas}{\gmenos}\right)^k,
\end{equation}
for all $\ell \geq 0$, $\lfloor x\rfloor$ denoting the integer part of $x$. 

Expanding $G(z)$ we obtain $s_{\ell}$ in terms of $a_{\ell}$,
\begin{equation}\label{eq:slaux}
s_{\ell}=(-1)^{\ell}\left[\frac{\gmas}{\gmenos}(s_0+\mu)a_{\ell-2}-(s_1+\mu)
a_{\ell-1}\right]-\mu,
\end{equation}
for $\ell \geq 2$, where $a_{\ell}$ can be evaluated either using \eqref{eq:al} or 
\eqref{eq:als}. In order to solve the system \eqref{eq:rec}, we have to impose 
$s_{L+1}=0$ 
for an ecosystem to have $L$ trophic levels. This provides a linear relation between 
$s_0$ and $s_1$ which, together with the first equation of 
\eqref{eq:rec}, forms a linear system that determines both $s_0$ and $s_1$. The result is
\begin{equation}\label{eq:s0s1}
\begin{split}
s_0&=\frac{(R/n_c+\mu\gmenos+\mu)a_L-(-1)^L\mu\gmenos}{a_L+\gmas a_{L-1}}-\mu,\\
s_1&=\frac{\gmas(R/n_c+\mu\gmenos+\mu)a_{L-1}+(-1)^L\mu\gmenos}{\gmenos(a_L+
\gmas a_{L-1})}-\mu.
\end{split}
\end{equation}
Substituting \eqref{eq:s0s1} into \eqref{eq:slaux} and taking into account that
\begin{equation}\label{eq:sima}
a_L a_{\ell-2}-a_{L-1}a_{\ell-1}=(-1)^{\ell}\left(\frac{\gmas}{\gmenos}\right)^{\ell-1}
a_{L-\ell}
\end{equation}
is a direct consequence of the recurrence satisfied by $a_{\ell}$, we finally get
\begin{equation}\label{eq:sl}
\begin{split}
s_{\ell}&=\left(\frac{\gmas}{\gmenos}\right)^{\ell}\left(\frac{R}{n_c}+\mu\gmenos+\mu\right)
\frac{a_{L-\ell}}{a_L+\gmas a_{L-1}}\\
&-\mu\left[(-1)^{L+\ell}\frac{a_{\ell-1}+\gmas a_{\ell-2}}{a_L+\gmas a_{L-1}}+1\right]
\end{split}
\end{equation}
for all $\ell \geq 1$. This is
the analytic solution of the system \eqref{eq:rec} and gives an estimate
of the maximum occupancy per level as a function of the parameters of the model.
Note that, despite what \eqref{eq:al} might suggest, no additional factors of the form
$\gmas/\gmenos$ can be extracted from $a_{\ell}$ according to \eqref{eq:als}, so the
lowest power of the ratio $\gmas/\gmenos$ in the expression for $s_{\ell}$ is 
$(\gmas/\gmenos)^{\ell}$.

This dependence of $s_{\ell}$ on $(\gmas/\gmenos)^{\ell}$ is remarkable. In fact, in our 
previous 
work \citep{capitan:2009} we observed that the communities in the end 
states of the assembly process
were pyramidal. This is, in turn, a consequence of the exhaustion of the species 
occupancy in each trophic level. Notice also that the estimation of the 
maximum number of species
that a community can host depends linearly on the resource saturation.
This linear dependence on $R$ was also observed in our previous work. 

Our estimation of the maximum occupancy of each trophic level also provides a condition for
the maximum number of trophic levels that a set of parameters allows. Imposing
$s_L \geq 1$ yields a condition for the allowance of $L$ trophic levels,
\begin{equation}\label{eq:maxL}
\begin{split}
\frac{R}{n_c}+\mu(\gmenos+1) &\geq \left(\frac{\gmenos}{\gmas}\right)^L
\left[(1+\mu)(a_L+\gmas a_{L-1})\right.\\
&+\left.\mu(a_{L-1}+\gmas a_{L-2})\right].
\end{split}
\end{equation}
Therefore we have a minimum value of the resource saturation for $L$ trophic levels to be
viable in a community. 

\begin{figure}
\begin{center}
\includegraphics[width=88mm,clip=true]{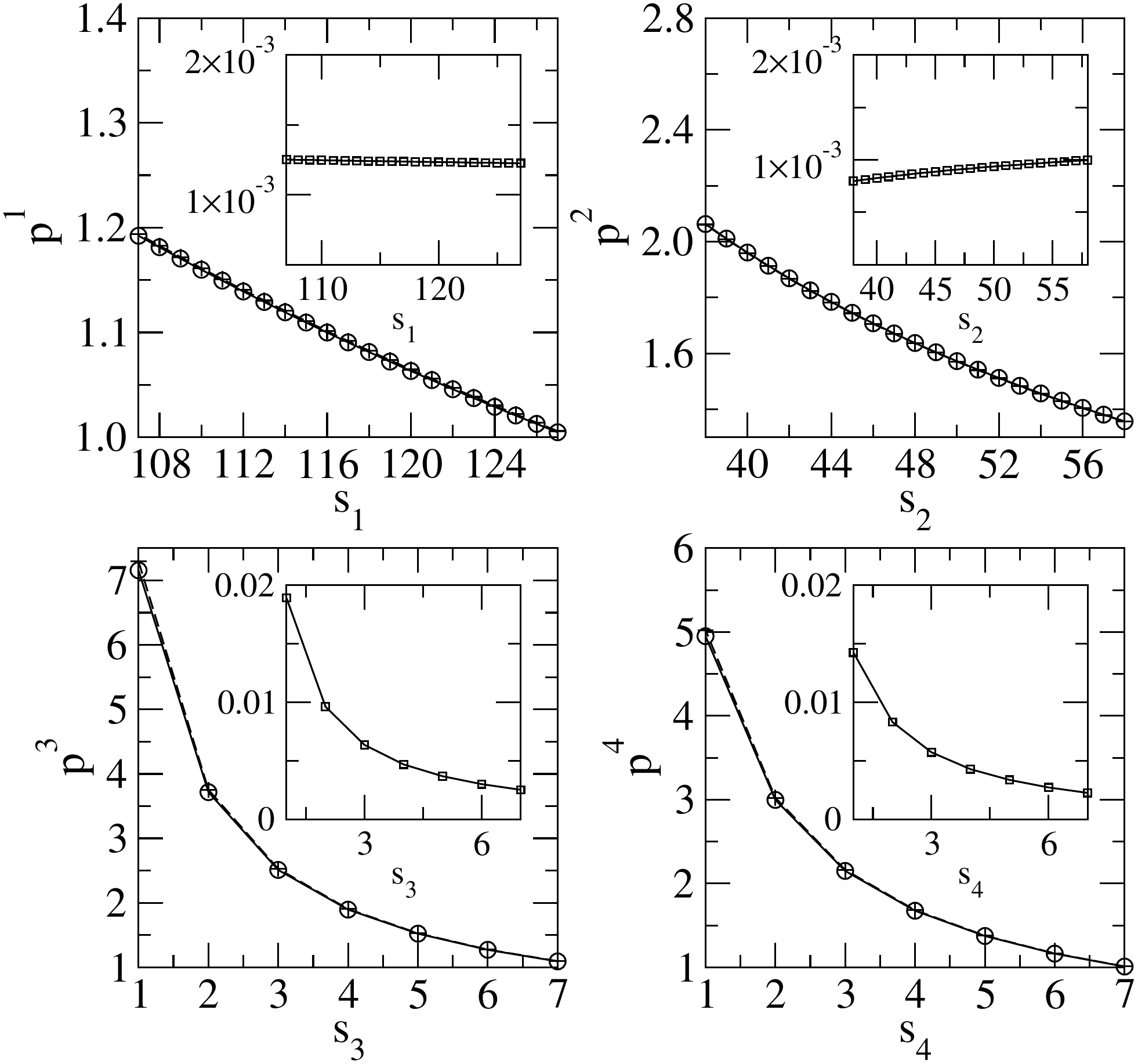}
\caption{Approximate equilibrium densities. Starting from a community
with 4 levels and occupancies $s_1=127$, $s_2=58$, $s_3=7$ and $s_4=7$,
we plot the variation of $p^{\ell}$ as a function of $s_{\ell}$, which exhibits a 
dependence $C/s_{\ell}$. Full lines with circles
show the exact solution of \eqref{eq:intsym}, and dotted lines with crosses show 
our approximation \eqref{eq:Plapp}. Insets contain the relative error of each
approximation. Remaining parameters are $R=1505$, $\gmas=0.5$, $\gmenos=5$,
$\rho=0.3$ and $\alpha=1$.}
\label{fig:eqdens}
\end{center}
\end{figure}

\subsection{Approximation of the equilibrium abundances}
\label{ss:nlsl}

In our model, each set $\{s_{\ell}\}_{\ell=0}^L$ of species occupancies in each level 
determines a set of equilibrium densities according to \eqref{eq:intsym}. Finding
$p^{\ell}(\{s_k\})$ is difficult, but in this section we will give 
a rather good approximation for large enough $s_{\ell}$. First we write the system 
in terms of the total population at each level, $P^{\ell}=s_{\ell}p^{\ell}$ 
($\ell=1,\dots, L$),
\begin{equation}\label{eq:intP}
\begin{split}
&\gmas P^{\ell-1}-\left(\rho+\frac{1-\rho}{s_{\ell}}\right)P^{\ell}
-\gmenos P^{\ell+1}=\alpha,\\
&P^0+\gmenos P^1=R.
\end{split}
\end{equation}
Written in this way, it seems natural to expand the solution in powers of $1/s$. 
In \ref{s:appB} we show that we can approximate
\begin{equation}\label{eq:Plapp}
P^{\ell} \approx \frac{T_{L,\ell}-(1-\rho)\sum_{k\neq \ell}^L 
Q_{L,\ell}^k/s_k}{D_L-(1-\rho)\sum_{k=1}^L B_{L,k}/s_k}.
\end{equation}
As we can see in Figure \ref{fig:eqdens}, this first order approximation captures
accurately the variation of the equilibrium densities $p^{\ell}$ with 
$s_{\ell}$. Besides, we also obtain a very accurate approximation when we vary 
the number of species $s_j $ in levels other than $\ell$. Note that, even when
the occupancy of a level is small (lower panels of Figure \ref{fig:eqdens}), 
the approximation remains good.

In the limit $s_{\ell} \gg 1$ we obtain the dependence $p^{\ell} \approx C/s_{\ell}$,
which provides the general tendency observed in Figure \ref{fig:eqdens}. Moreover,
in the biologically relevant limit $R \gg \alpha$, and taking into account the explicit
expressions for $T_{L,\ell}$ and $D_L$ given in \ref{s:appB}, populations behave like
\begin{equation}\label{eq:plapp}
p^{\ell}\approx\frac{R}{s_{\ell}}\left(\frac{\gmas}{\gmenos}\right)^{\ell}
\frac{a_{L-\ell}}{a_L+\gmas a_{L-1}}
\end{equation}
for $\ell \ge 0$. Several conclusions can be extracted from this dependence. First,
when the number of species in the $\ell$-th level is exhausted, according to \eqref{eq:sl}, 
we obtain a population density $p^{\ell} \approx n_c$, as expected. But more importantly, 
it represents another reason for the extinction threshold to be included in our model. 
If there were no threshold, equilibrium densities would monotonically decrease 
with $s^{\ell}$ without ever becoming zero. The assembly graph would then contain infinitely many
communities thus becoming intractable.

\section{Invaded dynamics}
\label{s:invaded}

In \cite{capitan:2009} it was assumed that, during the assembly process, successional 
invasions occur and modify resident communities {\em at equilibrium}. 
There we made the hypothesis of the average time between
consecutive invasions being much longer than the typical dynamic time scale for the 
community to reach the equilibrium state. This is actually what is observed. 
In relation to the different time scales
between invasion and competition, invasion events may take place at the scale of
decades, long enough time for invaded communities to
stabilize [for example, the rate of new invasions in islands may be one 
every few year \citep{sax:2005}]. This assumption has also been made in previous 
papers like \cite{kokkoris:1999}, where authors assume that after
each invasion there is a re-organization of the community prior to a new invasion.
Specifically, they solve the dynamical system describing
the new community with the invader until reaching the carrying capacity. These new
densities are then used as initial values for the new systems resulting from the
next invasion [see details in \cite{kokkoris:1999}]. The same idea was applied
in the construction of our assembly model \citep{capitan:2009}.

We used a second hypothesis as well, namely that the population of the invader is small
(equal to the extinction threshold $n_c$). This is what is actually
found in real situations. It is a well established 
fact that colonizers rarely reach a new habitat in high numbers \citep{roughgarden:1974,
turelli:1981}. In theory, the probability of a small propagule to extend
is used as the invasibility criterion. In biological control, management of invasions
is based on looking for a small density of species in new areas \citep{liebhold:2003}.
In this case, theoretical and empirical work has taken advantage to predict conditions
of eradication based on density thresholds (Allee effects) and demographic 
stochasticity.

Therefore we can assume invaders arriving at some level of a community 
in equilibrium with a small abundance 
set equal to the extinction threshold. Under the species symmetry assumption, the dynamic
system $\dot{n}_i^{\ell}=n_i^{\ell}q_i^{\ell}$ given by the response function 
\eqref{eq:L-Vsym} applies as well for the invaded system,
with $N^{\ell} = \sum_{i=1}^{s_{\ell}}n_i^{\ell}+n$ and $n$ being the population density 
of the invader.
Therefore, once the equilibrium is reached after the invasion, the density of the invader
will equal $p^{\ell}$ (the density of the remaining species in that level), which can
be obtained by solving \eqref{eq:L-Vred} with an occupancy $s_{\ell}+1$ in the $\ell$-th
level. Moreover, the global stability condition applies as well to the invaded dynamics. 
So we just need to check the viability of the resulting equilibria in order to determine 
whether the invader is accepted.

If the invasion takes place at level $L+1$, the equation for the invader is simply
\begin{equation}\label{eq:invasor}
\frac{\dot{n}}{n} = -\alpha + \gmas s_L n^L - n,
\end{equation}
which in fact is the last equation of the system \eqref{eq:L-Vred} 
for a community of $L+1$
levels with occupancies $\{s_1,\dots,s_L,1\}$. Hence the global stability condition
still remains applicable 
and the invader will be accepted if the resulting equilibrium is viable.

The complexity of the assembly dynamics comes from the cases where some level in the 
invaded community falls below the extinction threshold. The approach we used in 
\cite{capitan:2009} to determine the sequence in which species go extinct until leading to a final 
viable ecosystem was the following: for the levels 
that fell below the extinction threshold once the equilibrium had been reached, 
we went back in their trajectory
to the point where the population of some species crossed the extinction level $n_c$
for the first time,
we eliminated one species from that level and restarted the dynamics from that point. 
In this paper we will propose an alternative way to determine that sequence based on
several criteria and analytical approximations that we will discuss below.

\subsection{Invasion criteria}
\label{ss:criteria}

Consider the general dynamical system $\dot{x}_i/x_i = q_i(x,x_I)$, $\dot{x}_I/x_I =
q_I(x,x_I)$ for an arbitrary community with $S$ species, 
where $x$ are the densities of the species in the resident community and
$x_I$ is the density of the invader. The establishment of a colonizer in systems of 
this kind depends 
crucially on the initial per-capita growth rate of the invader \citep{law:1996}. In
fact, the condition that must be satisfied for a new species to increase when rare is
\begin{equation}
\lim_{T\rightarrow \infty} \frac{1}{T}\int_0^T q_I(\hat{x}(t),x_I=0) dt > 0,
\end{equation}
i.e., the time average of the per-capita rate of increase of the invader is positive
when the species of the resident community remain under certain attractor $\hat{x}(t)$ 
of the dynamics. In our model, the only attractor is the interior rest point, so the
condition reduces to $q_I(p,0) > 0$, where $p$ is the rest point of the resident
community. Strictly speaking, our model has a non-zero extinction threshold, so this 
condition has to be replaced by $q_I(p,n_c) > 0$. Since we start from a resident
community initially at equilibrium and the invader initial density is $n_c$, this 
condition reduces to the initial per-capita growth rate of the invader.

The condition $q_I(p,n_c)>0$ can be used to
obtain criteria for the invasibility at each level.
For example, consider the initial growth rate of the 
invader when the invasion takes place at the level $L+1$ [Eq. \eqref{eq:invasor}]. 
The condition for this rate to be positive is
\begin{equation}\label{eq:critL+1}
p^L > \frac{\alpha+n_c}{\gmas s_L}.
\end{equation}
If this condition does not hold, the invader is the first species to go extinct because 
it starts at the extinction level and with a negative initial rate. In the end states,
the populations of the resident community are close to (but above) $n_c$ 
\citep{capitan:2009}, so the former condition provides the approximate bound
\begin{equation}\label{eq:sLrec}
s_L \geq \frac{\alpha+n_c}{\gmas n_c}.
\end{equation}
Even if the initial growth rate of the invader is positive, asymptotically the level $L+1$
may not be viable. If this happens, during the time when the population of the invader
is above $n_c$, extinctions may occur at lower levels. This situation explains the
accumulation of recurrent states that we observed in \cite{capitan:2009} when we
varied the resource saturation (see Section \ref{s:assembly}).

Invasions at levels $\ell \leq L$ are subject to similar conditions. For the initial growth rate of
the invader to be positive
\begin{equation}\label{eq:critl}
p^{\ell} > \frac{n_c}{1-\rho}
\end{equation}
must hold.
In general, an initially positive growth rate could lead to potential extinctions in the
remaining levels while the equilibrium density of the invader is above the threshold. 
But it could happen as well that the
invader extinguishes at equilibrium with some initial transient time above the extinction.
To estimate a condition for this to happen, let us assume that densities and occupancies 
are inversely proportional (see \eqref{eq:Plapp} and Figure \ref{fig:eqdens}). Then the 
equilibrium abundance of the invader is $s_{\ell}p^{\ell}/(s_{\ell}+1)$, therefore if
\begin{equation}\label{eq:critl1}
p^{\ell} < n_c\left(1+\frac{1}{s_{\ell}}\right)
\end{equation}
the invader goes extinct. This condition, together with \eqref{eq:critl}, leads to
\begin{equation}\label{eq:critl2}
s_{\ell} < \frac{1}{\rho}-1
\end{equation}
so below this bound, 
the invader initially grows but becomes extinct at equilibrium. We will use this 
condition to explain the appearance of some recurrent subsets for certain values
of $R$ (see Section~\ref{s:assembly}).

It would be nice, however, to have a systematic way to predict the sequence of extinctions
after an invasion has occurred. Based on our approximations for the equilibrium densities,
we can propose a way to sequentially remove species for invasions at lower levels.
Within the end states of our model, abundances are close to the
extinction threshold. Then \eqref{eq:sl} implies that communities are pyramidal, so lower 
levels are
highly occupied but higher levels contain a small number of species. Accordingly,
the increase of one species in a lower level has no 
significant effect in the equilibrium abundances of the community.
Therefore if a species goes extinct after an invasion in a low level,
it has to be the invader itself.

The extinction sequence for invasions in higher levels is not so easy
to predict. Nevertheless, changes in abundances upon increasing $s_{\ell}$
are larger the higher the level (Figure \ref{fig:eqdens}) so,
in case that several levels fall below the threshold, we can make the assumption 
that it is always 
the ``highest'' species the one that goes extinct first. This procedure
provides a certain sequence of
extinctions whose accuracy will be checked in Section \ref{s:assembly}.

The prediction of the sequence of extinctions can be complicated 
when a top predator invades if the
resource saturation values do not allow for $L+1$ levels. We have
devised global approximations to the dynamics in this case to predict the order of
extinctions without having to resort to the numerical integration of the system of
differential equations, as we did in \cite{capitan:2009}.

\subsection{Global approximations to the dynamics invaded by a top predator}
\label{ss:approx}

Our heuristic approximations to the time dynamics of the system \eqref{eq:L-Vred} when
an invader arrives at level $L+1$ are somehow inspired in the matching technique used 
to obtain analytic 
approximations to perturbed differential equations [see, for example \cite{bender:1984}].
First we calculate the equilibrium point $\{p^{\ell}\}_{\ell=0}^L$ by either solving
\eqref{eq:intsym} or using the approximations \eqref{eq:Plapp}. Then
we approximate $n^{L+1}(t)$ by the sum of its long-term  
dependence $n_{\rm lt}^{L+1}(t)$ (near equilibrium) plus a 
short-term behavior $n_{\rm st}^{L+1}(t)$. For the long term, a linear stability
analysis shows that the solution
exponentially decays towards the equilibrium point, so we will set
\begin{equation}\label{eq:nL+1app}
n_{\rm lt}^{L+1}(t) = p^{L+1}+e^{-\lambda t}[d_0 \cos(\omega t)+d_1 \sin(\omega t)]
\end{equation}
where the eigenvalue of the linear stability matrix which is closest to zero is
$-\lambda+i\omega$ ($\omega$ may be zero). The constants
$d_0$ and $d_1$ remain undetermined for the moment.

\begin{figure}
\begin{center}
\includegraphics[width=88mm,clip=true]{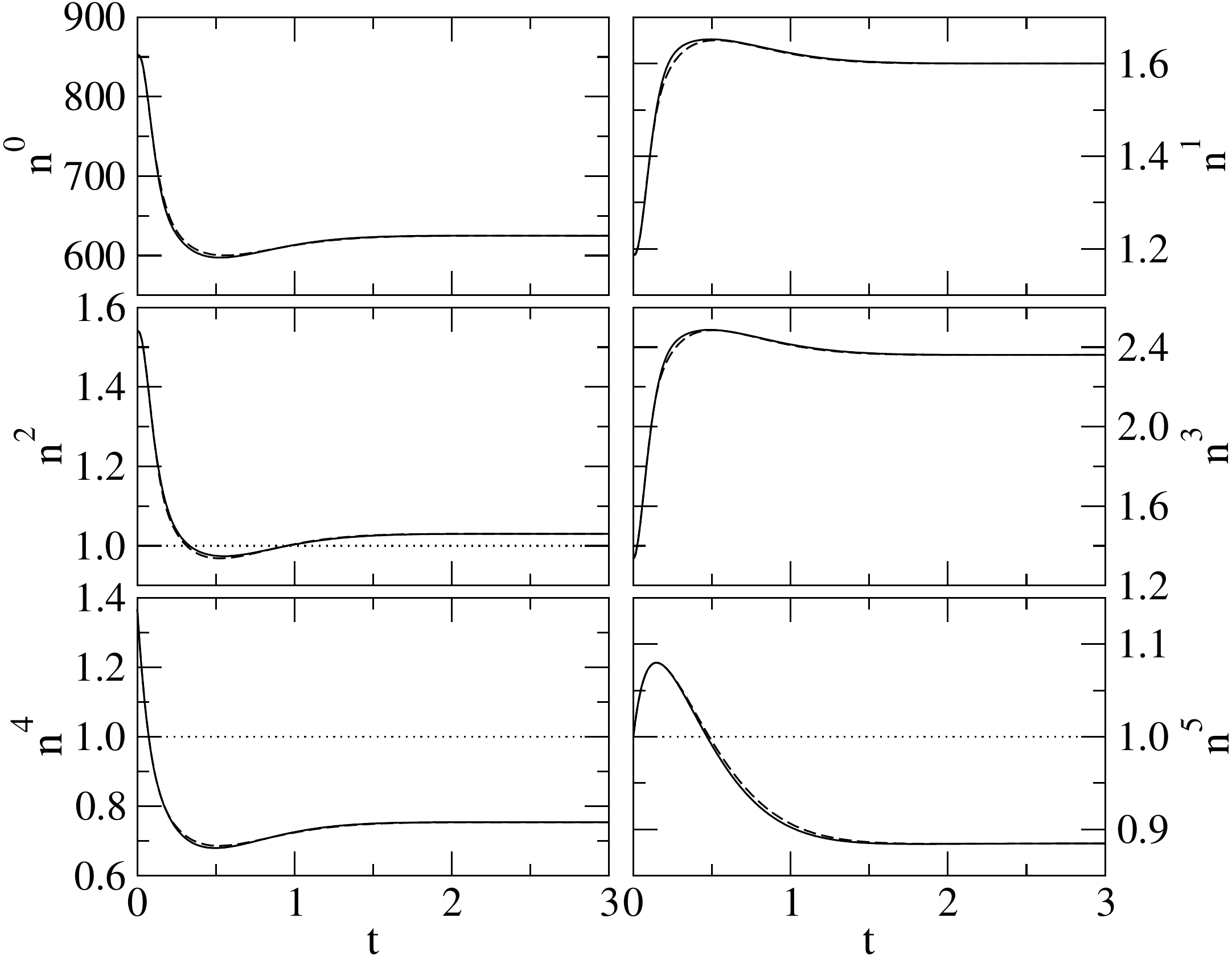}
\caption{Dashed lines show our approximation for the dynamics of a four-level
community determined by the occupancies $s_1=110$, $s_2=50$, $s_3=6$ and $s_4=5$
when invaded by a top predator at level $5$. 
For this case the eigenvalue closest to zero of the
linear stability matrix is complex. Full lines
represent the numerical integration of \eqref{eq:L-Vsym}. Remaining parameters are the 
same as in Figure \ref{fig:eqdens}. The whole time evolution is accurately predicted. The 
extinction level $n_c=1$ is showed as a dotted line. We can see how the first extinction
in the community takes place at level $4$.}
\label{fig:dynR1505}
\end{center}
\end{figure}

\begin{figure}
\begin{center}
\includegraphics[width=88mm,clip=true]{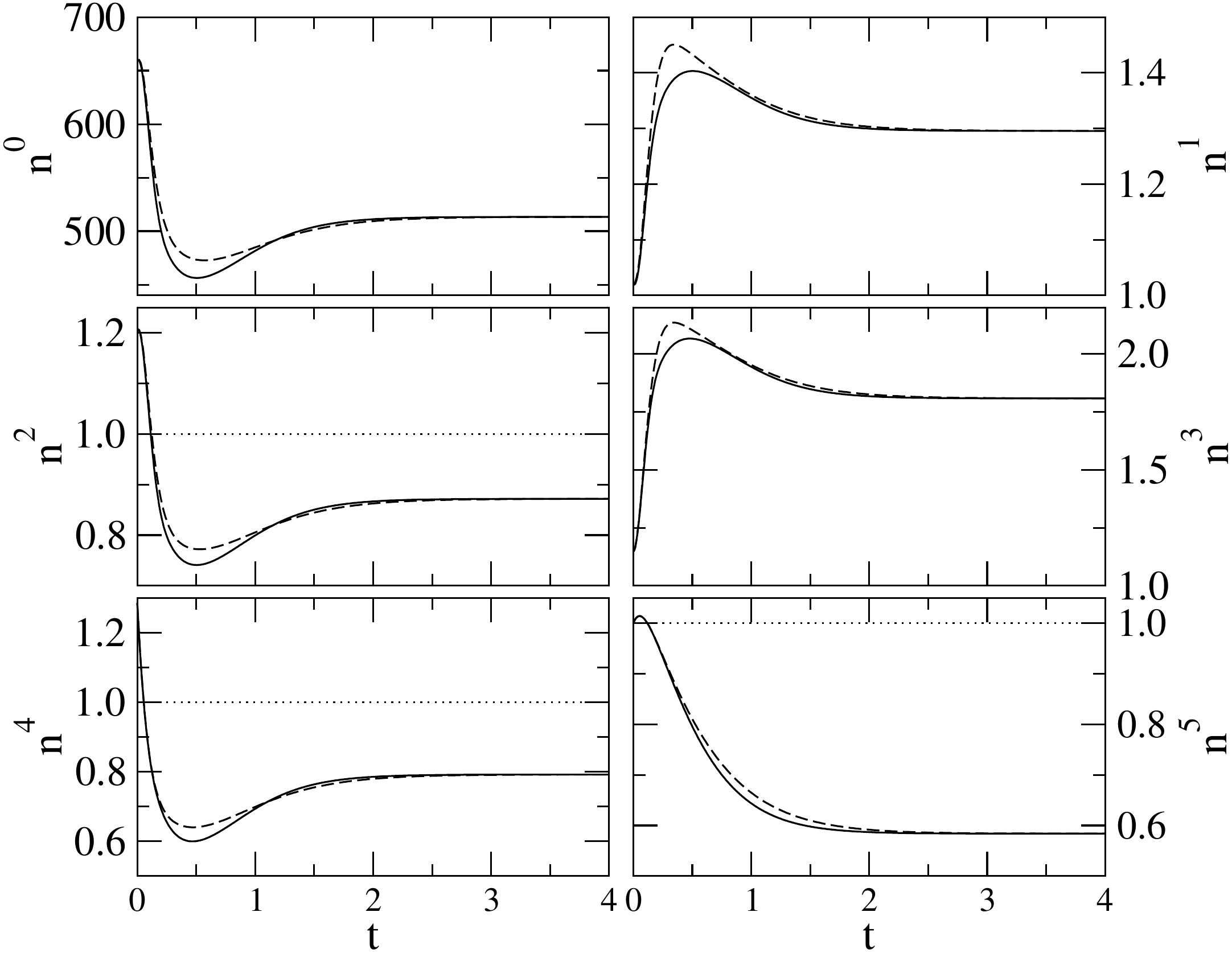}
\caption{Same as Figure \ref{fig:dynR1505}, but with $R=1200$ and occupancies 
$s_1=106$, $s_2=49$, $s_3=6$ and $s_4=4$. For this case the eigenvalue closest to zero of the
linear stability matrix is real. Although there is some discrepancy in our
approximations, the global trend is captured and the
extinction times after the invasion are accurately predicted.}
\label{fig:dynR1200}
\end{center}
\end{figure}

For the short-term behavior we propose
\begin{equation}\label{eq:nL+1short}
n_{\rm st}^{L+1}(t) = C(t)e^{-\xi t},
\end{equation}
where $C(t)=\sum_j c_j t^j$ is a polynomial whose coefficients and the exponent 
$\xi$ need to be determined to capture the transient time evolution. This way to express
the short-term behavior is inspired in the initial transient decay that can be observed
in the initial dynamics prior to getting close to the equilibrium point 
(see Figures \ref{fig:dynR1505} and \ref{fig:dynR1200}). The polynomial
has been included so as to properly capture the initial condition and the
initial deviations to the exponential decay. The technical details to calculate
the undetermined coefficients in \eqref{eq:nL+1app} and \eqref{eq:nL+1short}
are deferred to \ref{s:appC}. Figures~\ref{fig:dynR1505} and \ref{fig:dynR1200}
illustrate the validity of this approximation in capturing the global trend of
the time evolution. 

\begin{figure}
\begin{center}
\includegraphics[width=88mm,clip=true]{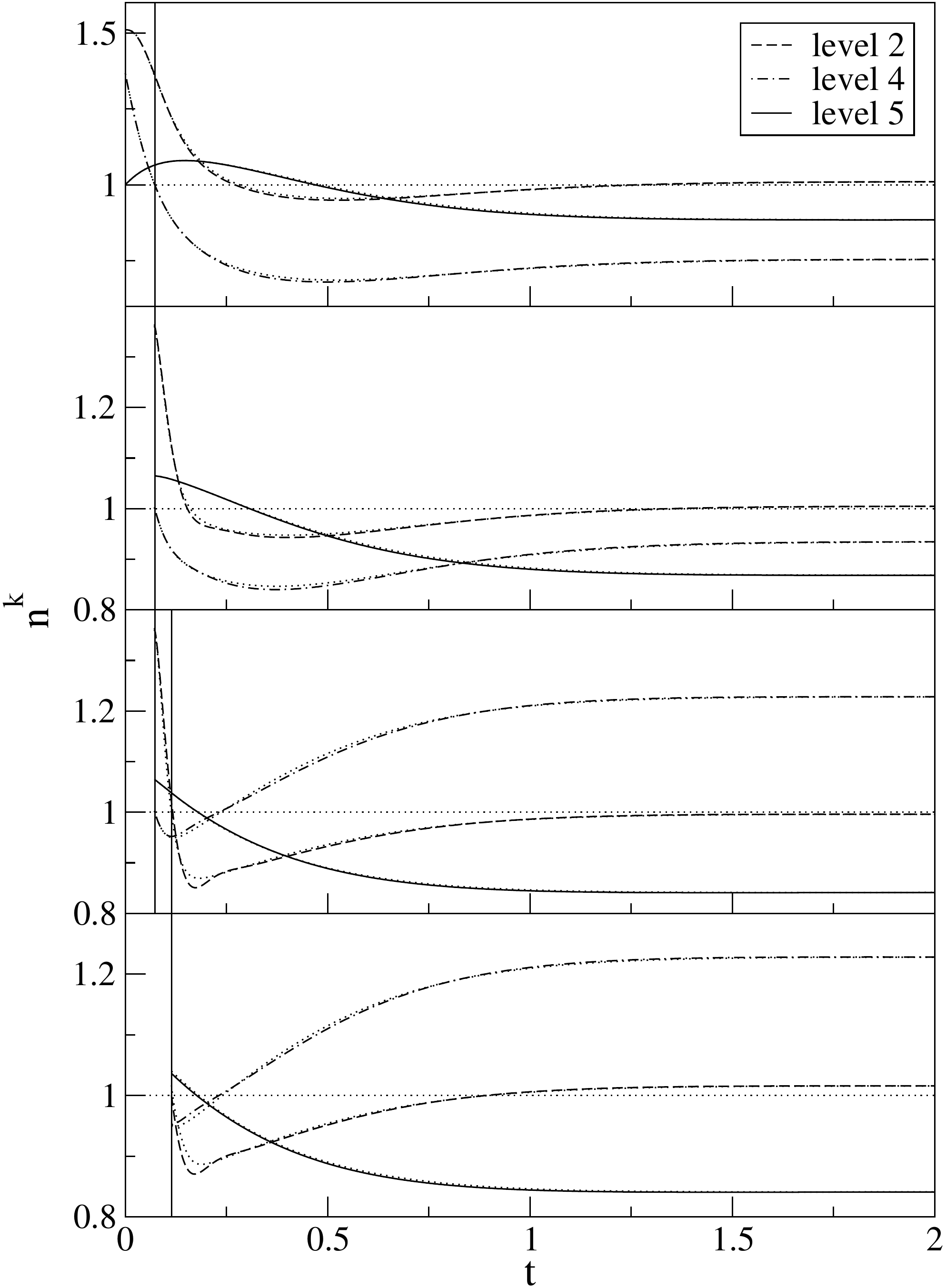}
\caption{Extinctions sequence for the community with $s_1=110$, 
$s_2=51$, $s_3=6$ and $s_4=5$ invaded at level 5 (parameter values are the same
as in Figure \ref{fig:eqdens}, and $n_c=1$ is showed with a horizontal dotted line). 
We just show the time evolution of the levels
that go extinct or are close to extinction in equilibrium. Dotted curves correspond
to our analytical approximations. We show, with vertical lines, the time
of the first level that go extinct. The sequence of extinct levels is 4, 4, 2, 5
until viability is recovered.}
\label{fig:extR1505}
\end{center}
\end{figure}

To reproduce the ordering of the extinctions we need the extinction times for each level, 
and these times are approximated with a higher accuracy than the dynamic trajectories
themselves (see Figure~\ref{fig:dynR1200}). In Figure
\ref{fig:extR1505} we illustrate, for a particular community, the extinction procedure 
compared to our analytical approximations. In this case, the first level
falling below $n_c$ is the fourth one (upper panel). Then we remove one species from
that level and restart the dynamics from the point of extinction, and the fourth
level falls again below $n_c$ (second panel). After the removal of a new species,
the fourth level ends up above $n_c$ at equilibrium. Now the next level ending
below $n_c$ is the second one. We
move to the point of extinction of this second level and restart the dynamics after
removing one species from $\ell=2$. After that it is just the invader ($\ell=5$) the only 
one that falls below the threshold, so we remove it and the resulting 
community becomes viable. Were it not, we would apply the same extinction procedure 
again and again until the final community is viable. The sequence of 
extinctions is well reproduced with our approximate solution, although slight
differences that alter the order of extinctions 
may occur when different levels fall below $n_c$ roughly at the same time.

\section{Application to community assembly}
\label{s:assembly}

Our goal in this paper was to provide analytical support,
albeit approximate, to the results obtained in
\cite{capitan:2009}. We want to check now whether our approximations correctly predict 
the recurrent sets which are end states
of the assembly process. With this aim, we have varied the parameter
$R$ within the range from 10 to 1700 in steps $\Delta R=5$.
The remaining parameters of the model will be set as in our previous work: $\gmas=0.5$,
$\gmenos=5$, $\rho=0.3$, $\alpha=1$ and $n_c=1$.

Let us first fix the number of levels $L$. We can determine with \eqref{eq:maxL} 
the minimum value $R_{\rm min}$ that allows $L+1$ levels. 
The results are summarized in Table \ref{tab:maxL}. Moreover, we can
combine \eqref{eq:sl} and \eqref{eq:sLrec} to give an estimation of the initial value
of $R_{\rm rec}$ for the appearance of a recurrent set with more than one community,
\begin{equation}\label{eq:minL}
\begin{split}
\frac{R}{n_c} +\mu(\gmenos+1) &\geq \left(\frac{\gmenos}{\gmas}\right)^L
\Biggl[\left(\frac{\alpha+n_c}{\gmas n_c}+\mu\right)(a_L+\gmas a_{L-1})\Biggr.\\
&+\Biggl.\mu(a_{L-1}+\gmas a_{L-2})\Biggr].
\end{split}
\end{equation}
The resulting values show a good agreement with those obtained numerically in 
\cite{capitan:2009} (see Table \ref{tab:maxL}).

\begin{table}
\begin{center}
\begin{tabular}{ccc}
\hline\hline
$L$ & $R_{\rm rec}/n_c$ & $R^*_{\rm rec}/n_c$ $(\pm 5)$\\
\hline
1 & 25.80 & $30$\\
2 & 75.88 & $80$ \\
3 & 323.93 & $325$\\
4 & 973.56 & $975$\\
\hline\hline
\end{tabular}
\begin{tabular}{ccc}
\hline\hline
$L$ & $R_{\rm min}/n_c$ & $R^*_{\rm min}/n_c$ $(\pm 5)$\\
\hline
2 & 35.80 & $40$ \\
3 & 131.88 & $135$\\
4 & 457.53 & $470$\\
5 & 1613.71 & $1630$\\
\hline\hline
\end{tabular}
\caption{ Estimation of the value of $R/n_c$ for the appearance of a
recurrent set with more than one community (left).
Minimum values of $R/n_c$ that allow a community with $L$ levels, according to 
\eqref{eq:maxL} (right). The interval of values of $R$ that correspond to the
recurrent sets is approximately $[R_{\rm rec},R_{\rm min}]$. $R^*_{\rm rec}$
and $R^*_{\rm min}$ are the corresponding values found using numerical
analysis \citep{capitan:2009} mapping the whole range of $R$ with a
resolution $\Delta R=5$.}
\label{tab:maxL}
\end{center}
\end{table}

\begin{figure}
\begin{center}
\includegraphics[width=80mm,clip=true]{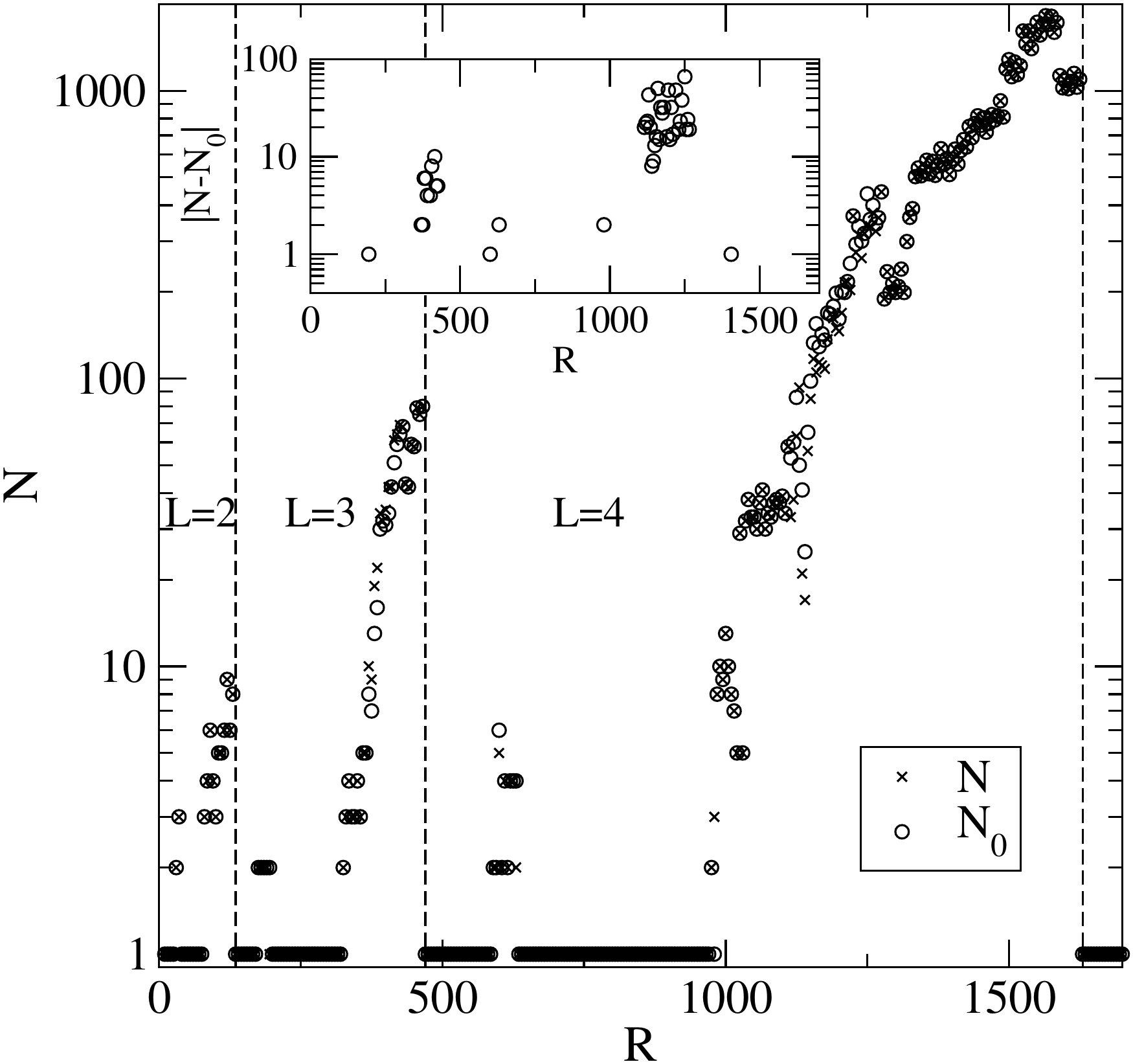}
\caption{Number of communities in the recurrent sets obtained with the analytical 
approximations ($N$, showed 
with crosses) and with a numerical integration of the population dynamics ($N_0$, circles).
The inset contains the absolute difference $|N-N_0|$. The global picture is the same as that
found in \cite{capitan:2009}, although differences of a few tens arise in some cases.}
\label{fig:reccom}
\end{center}
\end{figure}

\begin{figure}
\begin{center}
\includegraphics[width=80mm,clip=true]{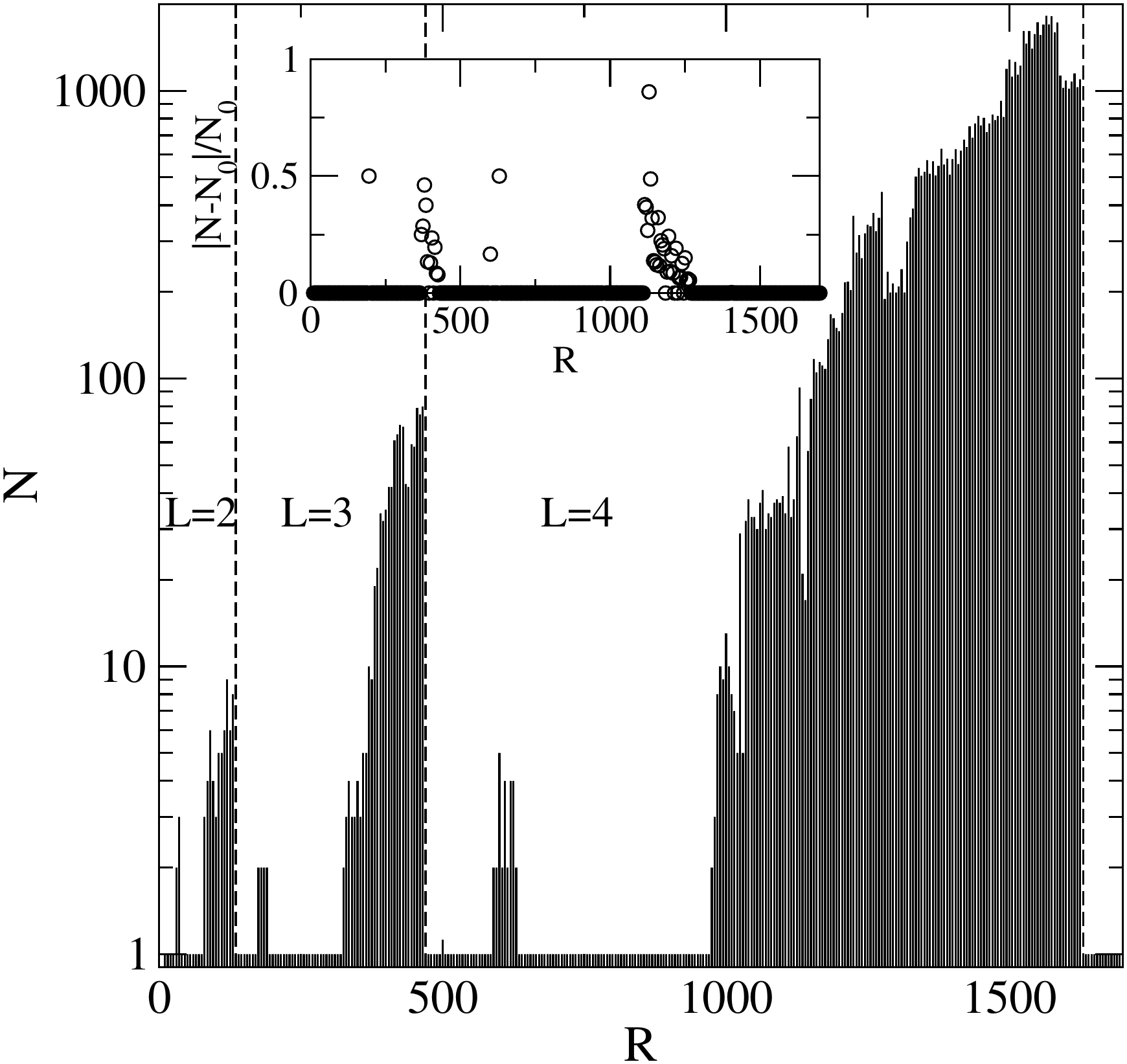}
\caption{Number $N$ of communities in the recurrent sets obtained with the analytical 
approximations (crosses in Figure \ref{fig:reccom}). The global trend is the same as
found in our previous work [see Figure 3. in \cite{capitan:2009}]. The inset shows the 
relative difference in the prediction of the number of communities in the end states.
Note that the discrepancies occur in a region where this number is small. This
explains the relatively large error found in some cases.}
\label{fig:anrec}
\end{center}
\end{figure}

Then, for a given $R$, we can read off from Table \ref{tab:maxL} the number
of levels for the communities within the recurrent set. Once we know it,
we determine with \eqref{eq:sl} an estimation for the maximum occupancies
allowed. We round off the estimates to get an integer set of values $\{s_{\ell}\}$ and
calculate the associated interior equilibrium. It can happen that some of the $p^{\ell}$
fall below $n_c$, so we decrease the corresponding occupancies $s_{\ell}$ eliminating
species one by one until the equilibrium turns out to be viable. 
This way we obtain a community very close to those of the recurrent set
(communities within this set are close to extinction), so we can use it as the 
initial community to start the assembly process. We then compute the set of
viable communities connected to it, which defines an assembly
graph much smaller than those obtained in \cite{capitan:2009} starting from
the empty community $\varnothing$. We analyze the graph to obtain its recurrent sets
using the algorithm of \cite{xie:1998} 
and we get one single set. In Figures \ref{fig:reccom} and \ref{fig:anrec} we plot the 
number of communities in each end state, showing a good agreement between the 
results obtained with the analytical approximations reported here 
and the numerical results reported in \cite{capitan:2009}.

For every $R$ we can always find a community which is uninvadable at all its 
levels $\ell\leq L$. If $R$ is such that \eqref{eq:minL} is not verified, 
then the invader at level $L+1$ initially decreases and goes extinct. 
This explains the intervals of $R$ where only one absorbent state is found. However,
if \eqref{eq:sLrec} holds (with our choice of parameters this happens when
$s_L \geq 4$), there is an initial time interval where the population of the
invader is above the threshold. This can cause the extinction of lower level
species, and generate recurrent sets with more than one community.

Our analytical approximations thus provide results very close to those obtained numerically.
Besides its being more efficient (the whole assembly needs not be generated), this method 
also allows to predict what would happen for values of $R$ larger
than 1700, which are computationally prohibitive for the numerical method.
With our bounds \eqref{eq:maxL} and \eqref{eq:minL} we can estimate the next
interval of $R$ where more than one community in the end state will appear,
namely $R \in [3844, 5114]$. That is out of reach of the numerical method, because
the number of communities in the whole assembly graph grows as fast as 
$N \approx e^{\kappa \sqrt{R}}$ \citep{capitan:2010b}.

Two observations are on purpose. First, there are small intervals of $R$
where the graph constructed starting from the empty community has $L$ levels but there
are viable communities with $L+1$ levels which cannot be assembled starting from
$\varnothing$ [this phenomenon is analogous to the existence of unreachable persistent
communities showed in \cite{warren:2003}]. We observe this for $R=$ 460, 465, 1615, 1620 
and 1625 (see Table~\ref{tab:maxL}). We have checked that even in these cases the recurrent
state is exactly recovered using the analytical approximations.

Secondly, we can observe from Figures \ref{fig:reccom} and \ref{fig:anrec} that there are
small regions where recurrent sets with more than one community are found out of
the intervals predicted in Table~\ref{tab:maxL} (around $R\approx 200$
for $L=3$ and $R \approx 620$ for $L=4$). For those values, a single absorbent community 
should be found. However, 
condition \eqref{eq:critl2} for an invader at level $L$ to initially grow and become
extinct at equilibrium renders $s_L \leq 2$ for our
choice of $\rho$. We have checked that this condition is satisfied by all 
these small recurrent sets, thus explaining their appearance.

\begin{figure}
\begin{center}
\includegraphics[width=75mm,clip=true]{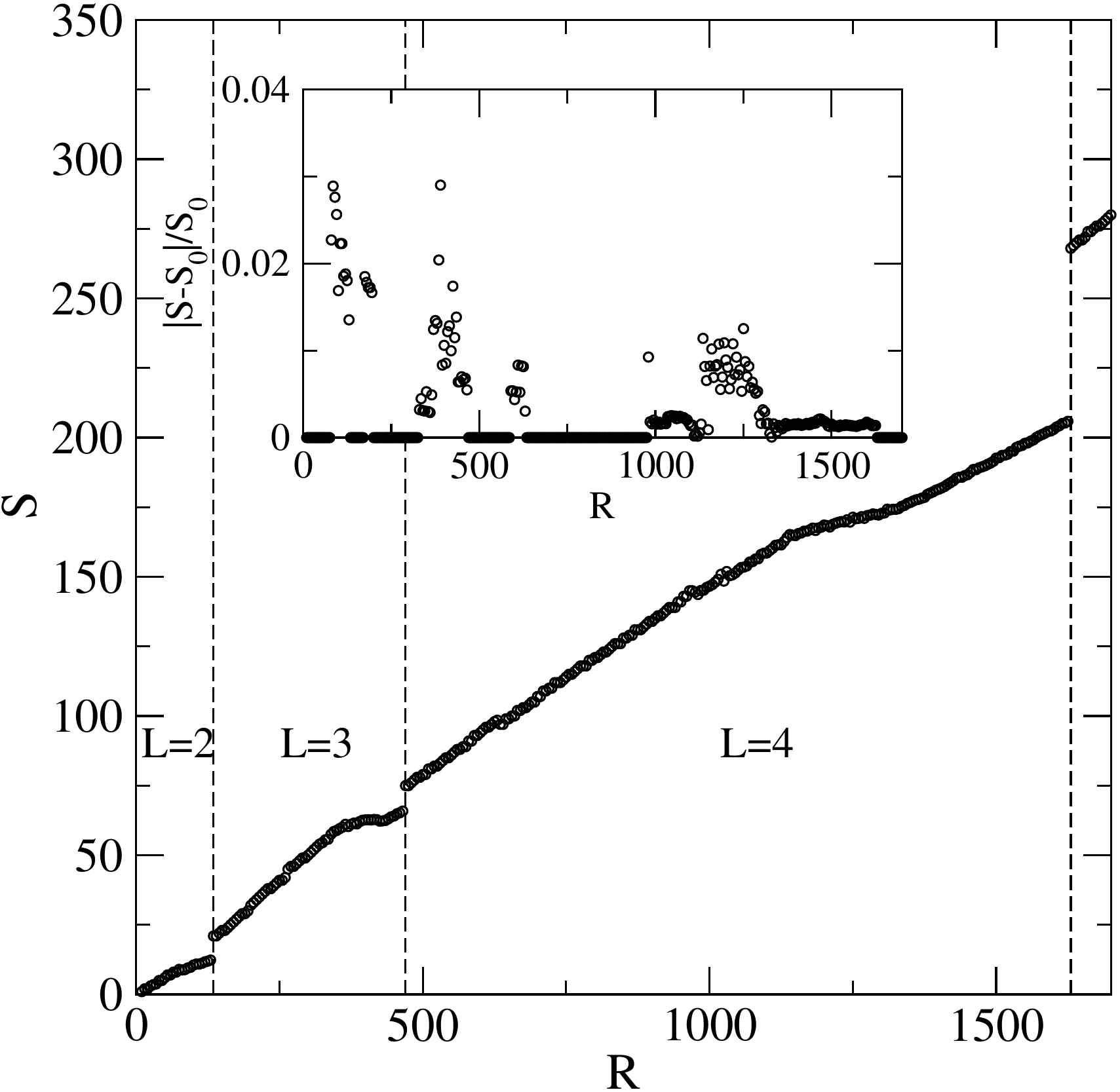}
\caption{Average number of species $S_{\rm av}$ in the end states calculated analytically 
vs. $R$. In the inset we show the relative error between $S$ and its corresponding 
average $S_0$ for each graph calculated numerically.}
\label{fig:avesp}
\end{center}
\end{figure}

We have to assess the accuracy of the transitions predicted in the graph of our recurrent
sets. Note that a slight difference in the ordering of extinctions can change the final
community after the invasion and this may change the observed graph and therefore
the asymptotic probability 
distribution of the associated Markov chain. In order to check the transition matrices
we obtain, we have calculated two averages. In Figure \ref{fig:avesp} we show the
variation of the average number of species in the recurrent sets as a function of $R$.
The behavior is almost indistinguishable from that found in \cite{capitan:2009} (the 
inset of Figure \ref{fig:avesp} shows that the relative error is small).

\begin{figure}
\begin{center}
\includegraphics[width=78mm,clip=true]{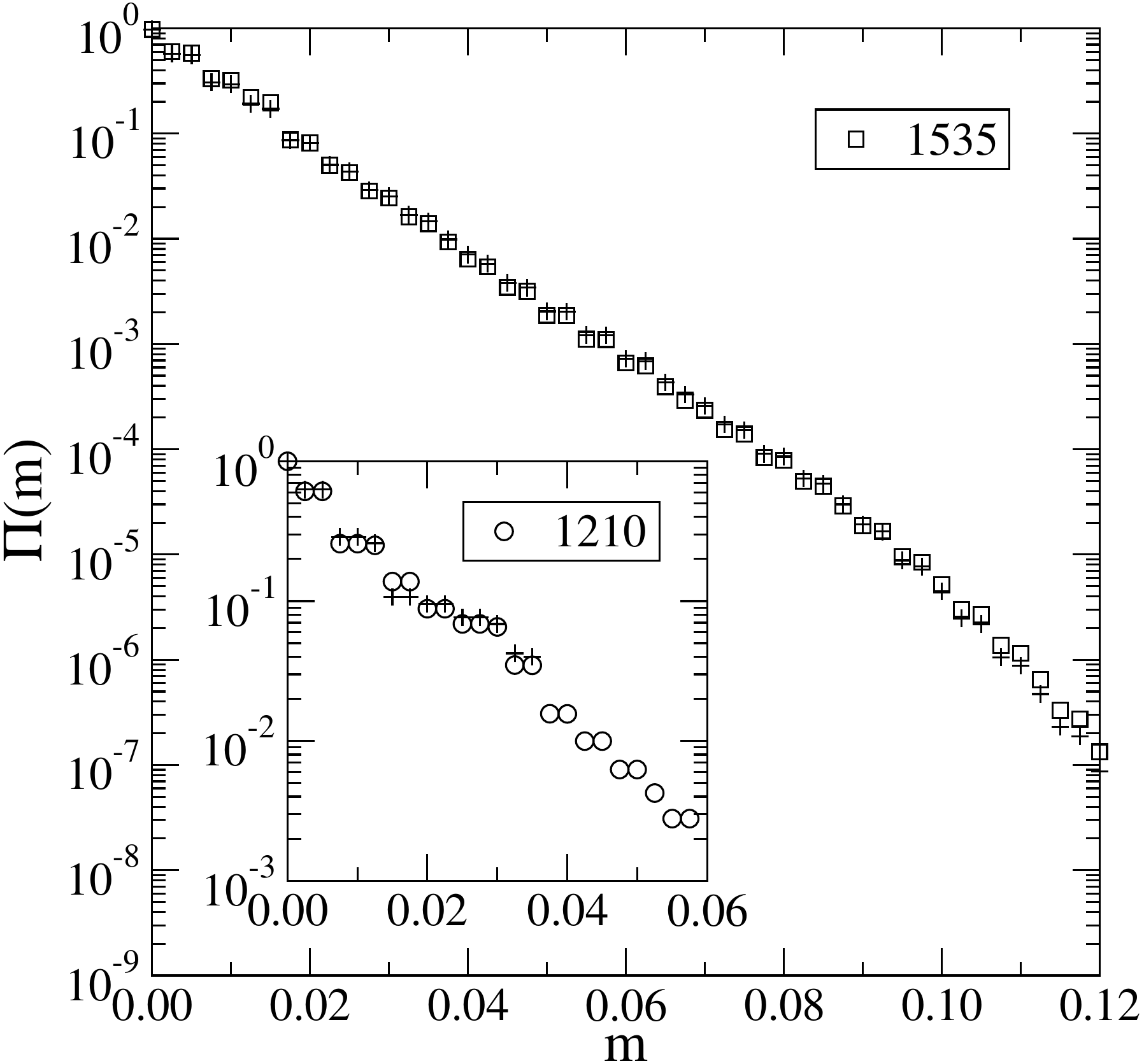}
\caption{Cumulative probability function $\Pi(m)$ for the distribution 
of the magnitude $m$ of avalanches of extinctions. The distributions follow an exponential
behavior. Crosses
represent the results for our approximated transition matrix. The number of recurrent states
coincide for the analytical and numerical method. The inset shows
a case where the number of communities is underestimated. This explains the absence of 
several points in the distribution estimated analytically. The agreement is rather good even 
in this case.}
\label{fig:aval}
\end{center}
\end{figure}

We have also checked that the number of extinctions predicted with our approximations
follows the same distribution than the one calculated numerically. To this purpose we
define the magnitude of an avalanche of extinctions as the relative variation $m=\Delta S/S$
of the total number of species in a community after an invasion. In Figure \ref{fig:aval}
we show the cumulative histogram for the distribution of these magnitudes. We can see that
the deviations between both distributions are small. Further 
statistical results will be discussed in the second paper of this suite.

\section{Conclusions}
\label{s:conclusions}

In this paper we have presented a general model of trophic-level structured food-web,
where interactions between species are either feeding or competing. For the sake of
simplicity, feeding only takes place between contiguous levels. The population
dynamics is modeled through Lotka-Volterra equations, and a proof is given that
a wide class of these models has a globally stable interior equilibrium. We have
introduced this model as an appropriate general framework to study the process
of successional invasions. In the invasion process, we consider a mean-field
version, in which species in the same level are trophically
equivalent and only intra- and interspecific competition is distinguished.
This species symmetry assumption has allowed us to obtain analytical results,
some of them exact and some other approximate. Among them we have provided
estimations for the maximum number of species allowed per level, the maximum
number of levels for a given value of the resource saturation, and certain
analytical approximations of the dependence of the equilibrium abundances on
the occupancies of each level. We have combined these results with some
criteria for the acceptance of an invader in our model communities, and with the help
of some global approximations of the invaded dynamics we have been able to
obtain, with high accuracy, the sequence of extinctions occurring after an invasion.
With this procedure we have reproduced the same results that we found in a
previous work \citep{capitan:2009}, this time without resorting to an integration of
the Lotka-Volterra equations and without constructing the whole assembly graph.
Among other things this brings the opportunity of exploring the model for
resources which would otherwise be computationally prohibitive to obtain.

Although the main results of this model are discussed at length in the second
paper of this suite \citep{capitan:2010b}, we have provided here a few of them
which illustrate the global assembly process and some of its main features. For
instance, we had reported already in \cite{capitan:2009} that, upon increasing
the resource saturation $R$, the number of levels, $L$, that the system is
able to sustain increases discontinuously. We provide here an estimate of
the values of $R$ at which this occurs, and show that this values grow
essentially as $\sim(\gmas/\gmenos)^L$. Under the assumption that populations
are close to the extinction level, we have shown that equilibrium communities
are pyramids ---again in agreement with the results obtained in 
\cite{capitan:2009}. Close to the onset of appearance of a new level, the
number of communities in the end state increases. We have identified that
the requirement for this to happen is that the population of a
top predator invading the community initially grows only to go eventually
extinct. From this knowledge we can estimate the value of $R$ at which
the end state starts to have more than just one community.

We have tested the approximations we have made by calculating some observables.
Among them we report on the average species richness as a function of $R$,
as well as the distribution of the avalanche of extinctions produced by
an invasion. In both cases the agreement is very good. In the latter case,
it is worth mentioning that this distribution of avalanches decays exponentially
with the avalanche size, meaning that there is a characteristic size
of the avalanches. This size roughly grows with the species richness of
the community, as one could expect. In any case, avalanches never get even
close to destroy the community.

We also propose in this paper an analytical approximation to the dynamics of 
a community invaded by a top predator. This approximation has been built
matching the initial behavior of the solution (derived from the initial
condition) and the asymptotic decay expected close to the equilibrium.
We have found a a rather good agreement with the solutions
obtained by a numerical integration of the Lotka-Volterra equations,
and has allowed us to correctly predict (in most of the cases) the
order of extinctions eventually caused by the invasion of a top predator.
These approximations have been applied to reproduce
the assembly graphs for the recurrent sets, showing small discrepancies
only for certain values of $R$. This provides an alternative method to
analyze the system for other sets of parameter values, with a negligible
computational cost compared to the construction of the whole assembly
graph.

Our assembly model is based on several
assumptions regarding the invasion process. Two of the most important ones
are that newcomers invade
at low population and the average time between invasions is large compared
to the time for the communities to reach the equilibrium. If the invasion
rate is too high \citep{fukami:2004,bastolla:2005b} or if the invasion
is not produced by rare species \citep{hewitt:2002}, the assembly process
---and hence the resulting end states--- can be drastically altered.
The reason is that communities that are not accessible from the equilibrium 
state may be so from a transient or if there is a massive invasion.
This changes the assembly graph in ways that we can neither predict nor even
check, because these processes are out of reach of our model. For instance,
considering invading transients, one of the strong simplifications we make
use of is that of starting always from a well-defined initial condition,
namely the equilibrium state. If the system can be invaded at any moment
during a transient there are infinitely many initial conditions to start
off from, something we cannot implement. So what happens if any 
of those two hypotheses is violated remains an open question.

\section{Acknowledgements}
This work is funded by projects MOSAICO, from Ministerio
de Educaci\'on y Ciencia (Spain) and MODELICO-CM, from Comunidad Aut\'onoma de Madrid
(Spain). The first author also acknowledges financial support through a contract
from Consejer\'{\i}a de Educaci\'on of Comunidad de Madrid and Fondo Social 
Europeo.

\appendix

\section{Derivation of the reduced dynamical system}
\label{s:appA}

We will show in this appendix that our dynamical system $\dot{n}_i^{\ell} = q_i^{\ell} 
n_i^{\ell}$, with the linear response function \eqref{eq:L-Vsym}, can be 
reduced to the form \eqref{eq:L-Vred} when all the initial species abundances at a certain 
level are equal. The crucial point for this to be true is the relation \eqref{eq:sym}.

This result can be formulated in a simple way. Consider the two-dimensional autonomous 
system
\begin{equation}
\begin{split}
\dot{x} &= f(x,y),\\
\dot{y} &= g(x,y),
\end{split}
\end{equation}
with the initial condition $x(0)=y(0)$ and which satisfies $f(x,y)=g(y,x)$. We are going to show 
that the Taylor expansions centered at $t=0$ of $x$ and $y$ are identical. In principle, 
both expansions will have certain radii of convergence. Let $t$ be lower than the minimum 
of these radii. Then we just need to show that all the derivatives at $t=0$ coincide. 
But this follows by induction.

The first derivatives are shown to be equal easily. Let us assume that 
$x^{(k)}(0)=y^{(k)}(0)$ for all $k=1,\dots,n$. Then the $(n+1)$-th derivative is
\begin{equation}
x^{(n+1)}(0)=\sum_{j=0}^n \left(\begin{array}{c}n\\j\end{array}\right)
\left.\frac{\partial^n f}{\partial x^j \partial y^{n-j}}\right|_{t=0} x^{(j)}(0) y^{(n-j)}(0).
\end{equation}
But, since $f(x,y)=g(y,x)$, this is equivalent to write
\begin{equation}
x^{(n+1)}(0)=\sum_{j=0}^n \left(\begin{array}{c}n\\j\end{array}\right)
\left.\frac{\partial^n g}{\partial y^j \partial x^{n-j}}\right|_{t=0} y^{(j)}(0) x^{(n-j)}(0),
\end{equation}
and, relabelling the sum index,
\begin{equation}
x^{(n+1)}(0)=\sum_{j=0}^n \left(\begin{array}{c}n\\n-j\end{array}\right)
\left.\frac{\partial^n g}{\partial x^j \partial y^{n-j}}\right|_{t=0} x^{(j)}(0)
y^{(n-j)}(0),
\end{equation}
which is equal to $y^{(n+1)}(0)$.

Therefore we have shown that the Taylor expansions of $x(t)$ and $y(t)$ coincide. This 
means that $x(t)=y(t)$ within the radius of convergence of the series. For larger times,
we can apply the same argument by analytic continuation (we choose some $t_0$ in
the interval or convergence as the centering point for a new Taylor expansion, and repeat
the argument). Hence we conclude that $x(t)=y(t)$ for all $t$.

Note that the same considerations apply to our system \eqref{eq:L-Vsym}, so we can 
reduce considerably the complexity of the system and solve \eqref{eq:L-Vred} instead.

\section{Analytical approximation to the equilibrium densities}
\label{s:appB}

This appendix is devoted to solve the linear system for the equilibrium densities 
\eqref{eq:intP}. The solution of this system can be obtained through Cramer's rule as
\begin{equation}\label{eq:Cramer}
s_{\ell}p^{\ell} = \frac{\Xi_{L,\ell}}{\Delta_L}
\end{equation}
for certain determinants $\Xi_{L,\ell}$ and $\Delta_L$. Our approximation is based in some 
recurrent equations that can be obtained for these determinants.

Let us start with the $(L+1)\times(L+1)$ determinant
\begin{equation}
\Delta_L = \begin{array}{|ccccc|}
-1 & -\gmenos & 0 & \cdots & 0 \\
\gmas & -d_1 & -\gmenos & \cdots & 0 \\
0 & \gmas & -d_2 & \cdots & 0 \\
\vdots & \vdots & \vdots & & \vdots \\
0 & 0 & 0 & \cdots & -d_L \end{array},
\end{equation}
where $d_{\ell} \equiv \rho+\frac{1-\rho}{s_{\ell}}$. Hence the densities depend on 
$\{s_{\ell}\}_{\ell=1}^L$ only through the inverse of all the possible products 
$s_{i_1}s_{i_2}\cdots s_{i_k}$, for some combination
$(i_1,i_2,\dots,i_k)$ of $k$ elements of the set $\{1,2,\dots, L\}$. In the recurrent sets 
we get the highest occupancy of species in each level allowed by the resource 
according to \eqref{eq:s0s1}--\eqref{eq:sl}, so we expect that a rather good approximation 
for the equilibrium densities amounts to neglecting orders higher than $1/s$. Hence
\begin{equation}\label{eq:DeltaL}
\Delta_L = D_L-(1-\rho)\sum_{\ell=1}^L \frac{B_{L,\ell}}{s_{\ell}}+\mathcal{O}
\left(\frac{1}{s^2}\right),
\end{equation}
where
\begin{equation}
D_L = \begin{array}{|ccccc|}
-1 & -\gmenos & 0 & \cdots & 0 \\
\gmas & -\rho & -\gmenos & \cdots & 0 \\
0 & \gmas & -\rho & \cdots & 0 \\
\vdots & \vdots & \vdots & & \vdots \\
0 & 0 & 0 & \cdots & -\rho \end{array}
\end{equation}
has dimension $(L+1)\times(L+1)$ and $B_{L,\ell}$ is the determinant obtained by
substituting the $\ell$-th column of $D_L$ by the column vector $u_{\ell}$
whose components are $u_{\ell,i} = \delta_{\ell,i}$ (for $i=0,1,\dots L$).

The determinant $D_{\ell}$ satisfies the recursion 
\begin{equation}\label{eq:recD}
D_{\ell} = -\rho D_{\ell-1}+\gmas\gmenos D_{\ell-2},
\end{equation}
where $\ell = 1,2,\dots L$, $D_0 = -1$ and $D_1 = \rho+\gmas\gmenos$. This relation can be 
easily solved using a generating function. On the 
other hand, it is easy to see that $B_{L,\ell} = D_{\ell-1}E_{L-\ell-1}$, with $E_{\ell}$ 
the $(\ell+1)\times(\ell+1)$ determinant
\begin{equation}
E_{\ell} = \begin{array}{|ccccc|}
-\rho & -\gmenos & 0 & \cdots & 0 \\
\gmas & -\rho & -\gmenos & \cdots & 0 \\
0 & \gmas & -\rho & \cdots & 0 \\
\vdots & \vdots & \vdots & & \vdots \\
0 & 0 & 0 & \cdots & -\rho \end{array},
\end{equation}
which also satisfies recursion \eqref{eq:recD} with $E_0=-\rho$ and 
$E_1 = \rho^2+\gmas\gmenos$.

The generating function that results from \eqref{eq:recD} is
\begin{equation}
G(z)=\sum_{\ell=0}^{\infty} D_{\ell} z^{\ell} = 
\frac{D_0+(D_1+\rho D_0)z}{\gmas\gmenos z^2-\rho z -1},
\end{equation}
and after the series expansion we get
\begin{equation}
D_{\ell} = (-\gmenos)^{\ell-1}\left[(D_1+\rho D_0)a_{\ell-1}-\gmenos E_0 a_{\ell}\right],
\end{equation}
with $a_{\ell}$ given by \eqref{eq:al}. Then the following compact expressions
result
\begin{align}
D_{\ell} &= (-1)^{\ell+1}\gmenos^{\ell}\left[a_{\ell}+\gmas a_{\ell-1}\right],\\
E_{\ell} &= (-1)^{\ell+1}\gmenos^{\ell+1}a_{\ell+1}.
\end{align}

The explicit expression for $\Xi_{L,\ell}$ is obtained from $\Delta_L$ substituting
its $\ell$-th column by the $(L+1)\times 1$ column vector 
$(-R,\alpha,\dots,\alpha)^{\rm T}$. We can expand it up to leading order in powers of 
$1/s$ to get
\begin{equation}\label{eq:XiL}
\Xi_{L,\ell} = T_{L,\ell}-(1-\rho)\sum_{\substack {j=1\\ j\neq \ell}}^L 
\frac{Q_{L,\ell}^j}{s_j}+\mathcal{O}\left(\frac{1}{s^2}\right).
\end{equation}
where 
\begin{align}
T_{L,\ell} &=
\begin{array}{|ccccccc|}
-1 & -\gmenos & 0 & \cdots & -R & \cdots & 0 \\
\gmas & -\rho & -\gmenos & \cdots & \alpha & \cdots & 0 \\
0 & \gmas & -\rho & \cdots & \alpha & \cdots & 0 \\
\vdots & \vdots & \vdots & & \vdots & & \vdots \\
0 & 0 & 0 & \cdots & \alpha & \cdots & -\rho 
\end{array}\\
&\hspace*{4.2mm}\begin{array}{ccccccc}
(0) & \phantom{-\gmenos} & \phantom{-\gmenos} & \phantom{\,\cdots} & (\ell) & 
\phantom{\,\cdots} & (L) \\
\end{array}\nonumber
\end{align}
and $Q_{L,\ell}^j$ is the determinant that results when we substitute the $j$-th column 
of $T_{L,\ell}$ by $u_j$ ($j \neq \ell$).

Expanding $T_{L,\ell}$ along its first row we get
\begin{equation}\label{eq:PL}
T_{L,\ell} = -\alpha A_{L,\ell}+\alpha\gmas\gmenos A_{L-1,\ell-1}+(-1)^{\ell+1}R 
\gmas^{\ell} E_{L-\ell-1},
\end{equation}
where we define the new $i\times i$ determinants $A_{i,j}$ as
\begin{align}
A_{i,j} &= 
\begin{array}{|ccccccc|}
-\rho & -\gmenos & 0 & \cdots & 1 & \cdots & 0 \\
\gmas & -\rho & -\gmenos & \cdots & 1 & \cdots & 0 \\
0 & \gmas & -\rho & \cdots & 1 & \cdots & 0 \\
\vdots & \vdots & \vdots & & \vdots & & \vdots \\
0 & 0 & 0 & \cdots & 1 & \cdots & -\rho 
\end{array}\\
&\hspace*{4.2mm}\begin{array}{ccccccc}
(1) & \phantom{-\gmenos} & \phantom{-\gmenos} & \phantom{\cdots} & (j) & 
\phantom{\cdots} & (i) \\
\end{array}\nonumber
\end{align}
that satisfy the recurrence equation
\begin{equation}\label{eq:recAj}
A_{n,j} = -\rho A_{n-1,j}+\gmas\gmenos A_{n-2,j}+\gmenos^{n-j} E_{j-2},
\end{equation}
for $j = 1, 2, \dots, n-1$ (with the boundary conditions $A_{j,j+1}=0$ and $A_{j,0}=0$), and
\begin{equation}\label{eq:recAn}
A_{n,n} = -\gmas A_{n-1,n-1}+E_{n-2}.
\end{equation}
These relations can be explicitly solved. On the one hand, by definition $A_{1,1} = 1$, which 
amounts to choosing $E_{-1}\equiv1$ for this to be compatible with \eqref{eq:recAn}. 
Moreover, making use again of a generating function, the solution of \eqref{eq:recAn} is
\begin{equation}\label{eq:Aj}
A_{j,j} = \frac{(-1)^{j-1}\gmenos^{j}}{\gmenos-\gmas+\rho}\left[\frac{\rho}{\gmenos} 
a_{j-1}+\frac{\gmas+\rho}{\gmenos}a_{j-2}+\frac{\gmas}{\gmenos} a_{j-3}
-\left(\frac{\gmas}{\gmenos}\right)^j\right],
\end{equation}
for $j \geq 2$. On the other hand, the explicit solution of \eqref{eq:recAj} is
\begin{equation}\label{eq:Ajn}
\begin{split}
A_{j+k,j} &= (-1)^k\gmenos^{k+1} a_{k+1}A_{j,j}\\
&+\frac{\gmenos^k E_{j-2}}{\gmenos-\gmas+\rho}
\left[(-1)^{k+1} (\gmenos a_k-\gmas a_{k-1})+\gmenos\right],
\end{split}
\end{equation}
for $k \geq 1$. Therefore equations \eqref{eq:Aj} and \eqref{eq:Ajn}, together
with \eqref{eq:PL}, provide an explicit solution for the determinants $T_{L,\ell}$. 

Fortunately, $Q_{L,\ell}^j$ can be written in terms of the previous determinants since 
$Q_{L,\ell}$ is a block-diagonal determinant with two blocks that satisfies
\begin{align}
Q_{L,\ell}^j &= D_{j-1} A_{L-\ell,\ell-j},\quad \textrm{for} \quad k < j,\\
Q_{L,\ell}^j &= E_{L-j-1} T_{j-1,\ell},\quad\, \textrm{for} \quad k > j.
\end{align}
This completes the analytical approximation of the equilibrium densities of our dynamical 
model. We have derived explicit expressions for all the terms involved in \eqref{eq:Cramer}, 
\eqref{eq:DeltaL} and \eqref{eq:XiL} up to leading order in $1/s$. Moreover,
note that the same technique applied to find this approximation can be extended to obtain 
the exact dependence on $\{s_{\ell}\}_{\ell=1}^L$ of the abundances. Higher-order terms 
in powers of $1/s$ introduce in the corresponding determinants several column vectors 
of the type of $u_{\ell}$ making each determinant to be block-diagonal involving 
$D_{\ell}$, $E_{\ell}$, $A_{i,j}$ or $T_{\ell,k}^j$, so that the general solution contains 
in each term a product of a certain combination of these determinants. This explicit 
expression can in fact be written, but it is too cumbersome. The approximations
here obtained are both sufficiently simple and accurate enough to capture the behavior 
of population densities in the communities of the recurrent sets.

\section{Technical details of the global approximations to the dynamics}
\label{s:appC}

In this appendix we will describe the calculation of the undetermined parameters 
of our ansatz \eqref{eq:nL+1app}--\eqref{eq:nL+1short} for the dynamics of system
invaded by a top predator. We impose that the initial condition and the first
$k$ derivatives at $t=0$ match the exact values, which can be readily calculated. 
Indeed, our system has the form $\dot{x}_i=-\alpha x_i+x_i f_i(x)$, where
$f_i(x)=\sum_j b_{ij}x_j$ is a linear function. Therefore we can recursively
calculate the $s+1$ initial derivative as
\begin{equation}\label{eq:deriv}
x^{(s+1)}_i(0)=-\alpha x^{(s)}_i(0)+
\sum_{j=0}^{s}\left(\begin{array}{c}
s\\j\end{array}\right) x^{(s-j)}_i(0)f_i(x^{(j)}(0)).
\end{equation}

For a real eigenvalue ($\omega=0$), we choose $C(t)$ [see \eqref{eq:nL+1short}] 
to be a polynomial of degree $k-2$, and for a complex one ($\omega\neq 0$)
we choose degree $k-3$, in order to compensate for the extra undetermined
coefficient in the long-term behavior in this case. Equating the approximate
solution to the initial condition and the first $k-1$ derivatives of our
ansatz to the exact values leads to a linear system for the undetermined coefficients.
The equation for the $k$-th derivative yields a polynomial equation for $\xi$, namely
\begin{equation}\label{eq:poly}
\sum_{j=0}^{k-2} \left(\begin{array}{c}
k-2\\j\end{array}\right) H_j \xi^{k-j-2} = (\lambda^2+\omega^2)p^{L+1},
\end{equation}
when $\omega=0$, where
\begin{equation}
H_j = (\lambda^2+\omega^2)n^{(j)}(0)+2\lambda n^{(j+1)}(0)+n^{(j+2)}(0)
\end{equation}
and $n^{(j)}$ stands for the $j$-th derivative of $n^{L+1}$, which can be
calculated exactly using \eqref{eq:deriv}. For $\omega=0$ Eq.~\eqref{eq:poly}
gets replaced by
\begin{equation}\label{eq:poly1}
\sum_{j=0}^{k-1} \left(\begin{array}{c}
k-1\\j\end{array}\right) \left[\lambda n^{(j)}(0)+n^{(j+1)}(0)\right]
\xi^{k-j-1} = \lambda p^{L+1}.
\end{equation}
Afterwards, we just need to calculate the coefficients
$c_j$ and $d_0$ (and $d_1$, if $\omega\neq 0$) by solving the linear system that
they satisfy.

Once we have the approximate time behavior for $n^{L+1}$ we calculate analytically the
remaining populations $n^{\ell}$ by direct substitution into the system \eqref{eq:L-Vred},
taking advantage of the recursive form of these equations, once $n^{L+1}$ is known.
Notice that, since we have to calculate successive derivatives in order to get any
lower population, the accuracy of $n^{L+1}$ at short times degrades as we calculate
lower level populations. Fortunately the model produces communities with a small
number of trophic levels \citep{capitan:2009}. The choice $k=5$ seems to 
be enough to account for the dynamics of any community of up to $L=4$
levels invaded by a top predator (see  Figures \ref{fig:dynR1505} and
\ref{fig:dynR1200}). For the description of the dynamics of communities with a 
higher number of levels we would need to choose polynomials of higher
degree in our ansatz.

A final caveat needs to be made with respect to the calculation of $\xi$. We need it
to be positive, otherwise \eqref{eq:nL+1short} would be meaningless. Among all the
roots of \eqref{eq:poly} we choose the largest, positive, real solution, so that any 
possible initial oscillation of the polynomial $C(t)$ is damped by the exponential. 
In the majority of the dynamics that we have approximated (see
Section~\ref{s:assembly}), we are able to find a positive solution for $\xi$.
However, in some cases there is no positive solution. In those cases we just
minimize the difference between the exact $k$-th derivative and the approximate
one at $t=0$. This also produces an acceptable solution. In all minimization
procedures that we have run, a positive exponent $\xi$ is always found.

\bibliographystyle{model2-names}
\bibliography{ecology}







\end{document}